\documentclass[11pt]{article}
\usepackage{epsfig} 
\usepackage{amssymb} 
\usepackage{graphics}
\newcommand{\sect}[1]{ \section{#1} \setcounter{equation}{0} }

\newcommand{\pslash}{p \! \! \! /} 
 
\newcommand{\partialslash}{\partial \! \! \! /} 
\newcommand{\half}{\mbox{\small{$\frac{1}{2}$}}} 
\newcommand{\MSbar}{\overline{\mbox{MS}}} 
 
\newcommand{\Nc}{N_{\!c}}
\newcommand{\Nf}{N_{\!f}}
\newcommand{\NN}{{\cal N}} 
\newcommand{\OO}{{\cal O}} 
\newcommand{\pithree}{\mbox{\small{$\frac{\pi}{3}$}}}
\newcommand{\twopithree}{\mbox{\small{$\frac{2\pi}{3}$}}}

\newcommand{\nn}{\nonumber}
\newcommand{\order}{{\cal O}}
\newcommand{\nl}{\nn\\&&}

\newcommand{\sbx}{\scalebox{0.525}}
\newcommand{\defDiag}[2]{\expandafter\newcommand%
  \csname diag-#1\endcsname{#2}}
\newcommand{\diag}[1]{\csname diag-#1\endcsname}

\newcommand{\picj}[1]{\;\parbox[c]{40pt}{\begin{picture}(40,40)(0,0)
\SetWidth{1.0}\SetScale{1.0} #1 \end{picture}}\;}
\newcommand{\picjj}[1]{\parbox[c]{0pt}{\begin{picture}(0,40)(43,0)
\SetWidth{1.0}\SetScale{1.0} #1 \end{picture}}}

\newcommand{\picbj}[1]{\;\parbox[c]{60pt}{\begin{picture}(60,40)(0,0)
\SetWidth{1.0}\SetScale{1.0} #1 \end{picture}}\;}
\newcommand{\picbjj}[1]{\parbox[c]{0pt}{\begin{picture}(0,40)(63,0)
\SetWidth{1.0}\SetScale{1.0} #1 \end{picture}}}

\def\Asc(#1,#2)(#3,#4,#5){\CArc(#1,#2)(#3,#4,#5)}
\def\Lsc(#1,#2)(#3,#4){\Line(#1,#2)(#3,#4)}

\defDiag{1}{\sbx{\picj{
\Asc(20,20)(20,-90,270)}}}

\defDiag{7}{\sbx{\picj{
\Asc(20,20)(20,0,180) 
\Asc(20,20)(20,180,360)
\Lsc(40,20)(0,20)}}}

\defDiag{51}{\sbx{\picj{
\Asc(20,20)(20,0,180)
\Asc(20,2)(27,42,138) 
\Asc(20,38)(27,222,318)
\Asc(20,20)(20,-180,0)}}}

\defDiag{62}{\sbx{\!\!\picbj{
\Asc(35,20)(20,256,76) 
\Lsc(40,39.5)(20,39.5) 
\Asc(25,20)(20,104,284)
\Lsc(30,0)(20,39.5) 
\Lsc(40,39.5)(30,0)}\!\!}}

\defDiag{63}{\sbx{\picj{
\Asc(20,20)(20,-30,90)
\Asc(20,20)(20,90,210)
\Asc(20,20)(20,210,330)
\Lsc(3,10)(20,20) 
\Lsc(20,20)(20,40) 
\Lsc(37,10)(20,20)}}}

\defDiag{841}{\sbx{\picj{
\Asc(20,20)(20,0,180)
\Asc(20,30)(22.36,-153.43,-26.57)
\Lsc(0,20)(40,20)
\Asc(20,10)(22.36,26.57,153.43)
\Asc(20,20)(20,-180,0)}}}

\defDiag{993}{\sbx{\!\!\picbj{
\Asc(35,20)(20,256,346)
\Asc(35,20)(20,-14,76) 
\Lsc(40,39.5)(20,39.5) 
\Asc(25,20)(20,104,284)
\Lsc(30,0)(20,39.5) 
\Lsc(40,39.5)(30,0)
\Asc(38,23)(24,136,250)}\!\!}}

\defDiag{952}{\sbx{\picj{
\Asc(20,20)(20,90,210)
\Asc(20,20)(20,210,330) 
\Asc(20,20)(20,-30,90) 
\Lsc(3,10)(20,40)
\Lsc(37,10)(3,10) 
\Lsc(20,40)(37,10)}}}

\defDiag{1016}{\sbx{\picj{
\Asc(20,20)(20,0,90)
\Asc(20,20)(20,-180.,-90)
\Asc(40,0)(20,90,180)
\Asc(0,40)(20,-90,0)
\Asc(0,0)(20,0,90)
\Asc(20,20)(20,-90.,0)
\Asc(20,20)(20,90,180)}}}

\defDiag{1010}{\sbx{\picbj{
\Asc(20,20)(20,90,270)
\Asc(40,20)(20,-90,90) 
\Lsc(40,40)(20,40) 
\Lsc(20,0)(40,0)
\Lsc(20,0)(20,40) 
\Lsc(40,40)(40,0) 
\Lsc(20,40)(40,0)}}}

\defDiag{1009}{\sbx{\picbj{
\Asc(40,20)(20,-120,120)
\Asc(40,20)(20,120,180) 
\Asc(40,20)(20,180,240) 
\Asc(20,20)(20,60,300)
\Asc(20,20)(20,-60,0)
\Asc(20,20)(20,0,60) 
\Lsc(40,20)(20,20)}}}

\defDiag{1020}{\sbx{\!\!\picbj{
\Asc(35,20)(20,256,346)
\Asc(35,20)(20,-14,76) 
\Lsc(40,39.5)(20,39.5) 
\Asc(25,20)(20,104,284)
\Lsc(30,0)(20,39.5) 
\Lsc(40,39.5)(35,20) 
\Lsc(35,20)(30,0)
\Lsc(34,20)(53,15)}\!\!}}

\defDiag{1011}{\sbx{\picj{
\Asc(20,20)(20,90,180)
\Asc(20,20)(20,180,270) 
\Asc(20,20)(20,270,360) 
\Asc(20,20)(20,0,90)
\Lsc(20,20)(20,40) 
\Lsc(20,20)(20,0) 
\Lsc(0,20)(20,20) 
\Lsc(40,20)(20,20)}}}

\defDiag{1022}{\sbx{\picbj{
\Lsc(20,0)(40,0)
\Lsc(20,40)(40,40)
\Asc(20,20)(20,-180,-90)
\Asc(40,20)(20,-90,90)
\Lsc(20,20)(40,20)
\Lsc(20,0)(20,20)
\Lsc(40,20)(40,0)
\Lsc(20,20)(20,40)
\Lsc(40,20)(40,40)
\Asc(20,20)(20,90,180)}}}

\defDiag{511}{\sbx{\picbj{
\Asc(20,20)(20,90,270)
\Asc(40,20)(20,-90,90) 
\Lsc(40,40)(20,40) 
\Lsc(20,0)(40,0) 
\Lsc(20,0)(20,20)
\Lsc(20,20)(40,40) 
\Lsc(20,40)(28,33) 
\Lsc(33,28)(40,20) 
\Lsc(40,20)(40,0)
\Lsc(20,20)(40,20)}}}

\defDiag{841.1.3}{\diag{841}\sbx{\picjj{\Vertex(13,20){2.5} \Vertex(27,20){2.5}}}}
\defDiag{1009.1.2}{\diag{1009}\sbx{\picbjj{\Vertex(37,30){2.5}}}}
\defDiag{1011.1.2}{\diag{1011}\sbx{\picjj{\Vertex(6,34){2.5}}}}

\begin{document}

\mbox{}\hfill{LTH 1098, TTP16-034}
\vspace{0.7cm}

\begin{centering}


{\LARGE {\bf Four loop renormalization of the Gross-Neveu model}}

\vspace{1.2cm}
\noindent
{\bf J.A. Gracey}

\vspace{0.2cm}
\noindent
{Theoretical Physics Division, Department of Mathematical Sciences, University 
of Liverpool, P.O. Box 147, Liverpool, L69 3BX, United Kingdom} 

\vspace{0.3cm}
{\bf T. Luthe}

\vspace{0.2cm}
\noindent
{Institut f\"ur Theoretische Teilchenphysik, Karlsruhe Institute of Technology 
(KIT), Karlsruhe, Germany}

\vspace{0.3cm}
{\bf Y. Schr\"oder}

\vspace{0.2cm}
\noindent
{Grupo de Fisica de Altas Energias, Universidad del Bio-Bio, Casilla 447, 
Chillan, Chile}

\vspace{0.5cm}
\noindent
{} 

\end{centering}


\vspace{5cm} 
\noindent 
{\bf Abstract.} We renormalize the $SU(N)$ Gross-Neveu model in the modified 
minimal subtraction ($\MSbar$) scheme at four loops and determine the 
$\beta$-function at this order. The theory ceases to be multiplicatively 
renormalizable when dimensionally regularized due to the generation of 
evanescent $4$-fermi operators. The first of these appears at three loops and 
we correctly take their effect into account in deriving the renormalization 
group functions. We use the results to provide estimates of critical exponents
relevant to phase transitions in graphene.  

\newpage 

\sect{Introduction}

The Gross-Neveu model with an $O(N)$ or $SU(N)$ symmetry is a quantum field
theory of spin-$\half$ fields which interact quartically, \cite{1}. It is also 
known as the Ashkin-Teller model, \cite{2}. Aside from being perturbatively 
renormalizable in two dimensions rather than four it has many properties which 
are in common with Quantum Chromodynamics (QCD). For instance, it is 
asymptotically free and has dynamical symmetry breaking whereby the classically
massless fermions become massive in the true vacuum, \cite{1}. These together 
with other properties such as the existence of an exact $S$-matrix, \cite{3}, 
which provides the full bound state spectrum of the quantum theory, mean that 
the Gross-Neveu model has been used for many years as a laboratory to examine 
ideas which are harder to gain insight into in higher dimensional theories such
as QCD. 

The theory is also of interest due to its connections with problems in 
condensed matter theory. For example, it has been shown that in the so-called 
replica 
limit, $N$~$\to$~$0$, the $SU(N)$ theory describes the physics of the random 
bond Ising model, \cite{4}. See \cite{5}, for instance, for a review article. 
More recently, the Gross-Neveu model has been found to be connected to problems
in conformal field theory (CFT) in dimensions greater than two. Specifically it
lies in the same universality class at the Wilson-Fisher fixed point in 
$d$-dimensions as the Gross-Neveu-Yukawa theory whose critical dimension is 
four. 

Aside from the Ising model connection, \cite{4}, which has been explored 
recently in the CFT context using the conformal bootstrap programme, \cite{6},
the Gross-Neveu model has connections with recent developments in AdS/CFT 
theories. For example, see \cite{7} for a recent review on the background to 
this area. Given this particular relation to current problems in higher 
dimensional quantum field theories, it is worth noting that that analysis rests
on performing computations with the large $N$ or $1/N$ expansion. This 
expansion parameter, which is an alternative to conventional perturbation 
theory, is another feature which the Gross-Neveu model shares with QCD. In the 
latter case one can perform large $\Nc$ or large $\Nf$ expansions where $\Nc$ 
and $\Nf$ are the number of colours or (massless) quark flavours respectively. 
One advantage of an analysis using the large $N$ expansion in the Gross-Neveu 
model, or large $\Nf$ in the case of QCD, is that each theory is renormalizable
away from the critical dimension of either quantum field theory.
 
Given the centrality of the Gross-Neveu model to a wide range of applications
since its introduction, \cite{1}, it is worth noting that it has not been
renormalized in perturbation theory to as high a loop order as several of the
other basic theories such as $O(N)$ scalar $\phi^4$ theory or QCD itself. For 
instance, after the one loop results of \cite{1} the two loop renormalization
was carried out in \cite{8} and verified in \cite{9}. Subsequently, the three 
loop renormalization was performed in \cite{10} with the $\beta$-function
appearing independently in \cite{11,12}. At four loops the wave function and 
mass anomalous dimensions were determined in \cite{13}. However, from the point
of view of determining critical exponents in the replica limit, say, the 
information in these four loop terms is of no use until the $\beta$-function is 
known to that order too as it contains the location of the critical point to 
the same precision. This is the purpose of the article where we will complete 
the full four loop renormalization of the $SU(N)$ Gross-Neveu model by 
determining the coupling constant renormalization constant in the modified 
minimal subtraction ($\MSbar$) scheme. In terms of time lines this is a quarter
of a century since the three loop $\beta$-function of \cite{11,12}. While 
relatively long this is not dissimilar to the other basic theories mentioned 
already. For instance, the step from five, \cite{14,15}, to six loops, 
\cite{16,17,18}, for 
the scalar $\phi^4$ theory in four dimensions also took twenty-five years. 
Similarly from the appearance of the four loop $\MSbar$ QCD $\beta$-function,
\cite{19}, and its verification \cite{20}, until the recent $\MSbar$ evaluation
for $SU(3)$, \cite{21}, the time interval was nearly a score of years. The 
seven loop wave function renormalization of $\phi^4$ theory has also been 
determined recently with an indication that the $\beta$-function could be 
available soon outside the normal waiting period, \cite{17}. 

What is central to all these recent increases in the loop order of the 
$\beta$-functions has been the advance in computational technology. This does 
not solely mean computer hardware and speed. More crucial has been the 
creation, for instance, of an overall algorithm to handle the inordinate amount
of integration by parts of the exponentially large number of Feynman diagrams 
which occur at successive loop orders, \cite{22}. A consequence of this 
algorithm is the need to determine explicitly the value of a relatively small 
subset of integrals, termed masters, which cannot be computed using integration
by parts. One such approach has been to evaluate these numerically to very high
precision using difference relations, \cite{23}. Then analytic values can be 
adduced using the integer finding method of \cite{24} and a basis of 
transcendentals, such as multiple 
zeta values (MZV). Alternatively these difficult master integrals can in 
certain situations be determined directly by algebraic methods. For instance, 
high level mathematics, such as algebraic geometry and the development of the 
theory and properties of hyperlogarithms and MZVs, have crystallized into an 
algorithm such that all the parameter integrations in the Schwinger 
representation of certain master Feynman integrals can be found. For instance, 
this approach has been encoded in the {\sc Hyperint} package, \cite{25}. 

Given this 
background we have applied the latest machinery to tackle the four loop 
$\MSbar$ evaluation of the Gross-Neveu $\beta$-function. It transpires that 
this is not as straightforward as for the parallel computation in scalar field 
theories and to a lesser extent than for QCD. This is because in dimensionally 
regularizing the Gross-Neveu model, as is the usual regularization for 
multiloop renormalization of such theories, one loses multiplicative 
renormalizability, \cite{26,27,28}, which was observed in detail in 
\cite{29,30}. In essence, as with the treatment of $4$-fermi operators in 
effective field theories in four dimensions, evanescent operators are 
generated. These are operators which exist in $d$-dimensions but are absent
in the critical dimension of the actual field theory being renormalized. Their 
presence in the regularized theory cannot be ignored, as has been noted in the 
Gross-Neveu context, \cite{29,30,13}, since they have an effect on the 
determination of the renormalization group functions in the lifting of the 
regularization. However, in the Gross-Neveu model the effect of these 
evanescent operators on the renormalization group functions does not become 
manifest until four loops. This was recognized in \cite{28,29,30} and 
implemented in the construction of the four loop mass anomalous dimension in 
\cite{13}. Useful in this respect was the formalism developed to account for 
the effect the evanescence has on the renormalization group functions given in 
\cite{26,27}. Like scalar $\phi^4$ theory the wave function anomalous dimension
has no one loop term. So the effect of the evanescent operators will not be 
apparent before five loops for that particular renormalization group function. 
Given our interest in the four loop $\beta$-function here we will be careful in
computing the underlying $4$-point function which determines the coupling 
constant renormalization and in the same instance find the {\em new} evanescent 
operators which are generated at four loops. These will be required for any 
future {\em five} loop renormalization. 

The article is organized as follows. Section $2$ is devoted to reviewing the
formalism required to renormalize two dimensional theories with a $4$-fermi
interaction via the projection method of \cite{26,27}. The technical details of
how we evaluated the four loop Feynman graphs contributing to the 
renormalization of the $4$-point function are given in section $3$. The main
result for the $\beta$-function is given in section $4$ and applications to 
problems in condensed matter theory are given in section $5$. We provide brief
concluding remarks in section $6$. There are two appendices which respectively 
detail the tensor reduction construction and the numerical and analytic form of
the master integrals up to and including four loops.

\sect{Preliminaries}
\label{se:prelim}

We begin by summarizing the essential ingredients required to renormalize the
$SU(N)$ Gross-Neveu model to four loops in the $\MSbar$ scheme. The technical
details as to how the computations are performed will be devolved to a later
section. We have chosen to consider the $SU(N)$ theory rather than the $O(N)$ 
model because the former formulation has fewer terms in the vertex Feynman 
rule. Hence it minimizes the number of algebraic manipulations in the 
computations. The $O(N)$ renormalization group functions can be reconstructed 
from the final $SU(N)$ results. 

The bare $SU(N)$ Gross-Neveu Lagrangian is,
\cite{1}, 
\begin{equation}
L ~=~ i \bar{\psi}_0^{i} \partialslash \psi^{i}_0 ~-~ m_0 \bar{\psi}^{i}_0
\psi^{i}_0 ~+~ \frac{1}{2} g_0 ( \bar{\psi}^i_0 \psi^i_0 )^2
\label{lagbare}
\end{equation}
where we use ${}_0$ to denote a bare field or parameter and $g$ is the coupling
constant. In general terms (\ref{lagbare}) is renormalizable in two dimensions,
\cite{1}. We have included a mass for the fermion here partly to be complete
but also because we wish to avoid potential infrared issues when we come to
computing the relevant Feynman diagrams. When one takes traces over 
$\gamma$-matrices the Feynman integrals will have propagators similar to those 
of a bosonic field. It is well known that in two dimensions a bosonic
propagator is infrared divergent. So including a mass for the fermion will
ensure that all emerging divergences are ultraviolet in nature. The technical 
problem which arises is that when one considers the theory in higher dimensions
it ceases to be renormalizable since the interaction produces a dimensionful 
coupling constant. This observation has implications when one dimensionally 
regularizes (\ref{lagbare}) to initiate its renormalization. Specifically 
(\ref{lagbare}) ceases to be {\em multiplicatively} renormalizable, 
\cite{26,27}. Instead evanescent operators are generated in $d$-dimensions 
whose presence affects the derivation of the true renormalization group 
functions. This has been recognized in the earlier work of 
\cite{26,27,28,29,30}. Moreover, a procedure has been developed to account for 
the effect of these evanescent operators, \cite{26,27}, which we will use and 
extend to the case of the $\beta$-function computation here. To appreciate the 
issue in more depth it is instructive to recall the properties of the 
$\gamma$-algebra. In strictly integer dimensions the $\gamma$-matrices satisfy 
the Clifford algebra
\begin{equation}
\{ \gamma^\mu,\gamma^\nu \} ~=~ 2 \eta^{\mu\nu} ~.
\end{equation}
However when the spacetime dimension $d$ becomes a continuous variable the
matrices $\gamma^\mu$ cease to span the spinor space. Instead the basis of
$\gamma$-matrices needs to be expanded to a new set of matrices denoted by
$\Gamma^{\mu_1\ldots\mu_n}_{(n)}$ for all integers $n$~$\geq$~$0$. We have 
chosen to use the basis and definition of \cite{26,30,31,32,33} and they are 
defined by
\begin{equation}
\Gamma_{(n)}^{\mu_1 \mu_2 \ldots \mu_n} ~=~ \gamma^{[\mu_1} \gamma^{\mu_2}
\ldots \gamma^{\mu_n]}
\end{equation}
where a factor of $1/n!$ is understood within the total antisymmetrization on 
the right hand side. To clarify when $d$ is an integer dimension $D$, say, then
the basis of $\Gamma$-matrices is finite. This is because 
$\Gamma_{(n)}^{\mu_1 \mu_2 \ldots \mu_n}$~$=$~$0$ for $n$~$>$~$D$ due to the
antisymmetrization. Although we are considering the non-chiral Gross-Neveu
model, (\ref{lagbare}), it is worth noting that 
$\Gamma_{(5)}^{\mu_1 \mu_2 \mu_3 \mu_4 \mu_5}$ has no relation to the usual
$\gamma^5$ matrix. The renormalization of the chiral Gross-Neveu model is more
involved, \cite{26,27}, and beyond our present considerations. 
 
With these generalized $\gamma$-matrices the dimensionally regularized
Gross-Neveu model can be regarded as a special case of the more general
$d$-dimensional Lagrangian, \cite{26,27}, 
\begin{equation}
L ~=~ i \bar{\psi}_0^{i} \partialslash \psi^{i}_0 ~-~ m_0 \bar{\psi}^{i}_0
\psi^{i}_0 ~+~ \frac{1}{2} \sum_{n=0}^\infty g_{(n) \, 0} \, \bar{\psi}^i_0
\Gamma_{(n)}^{\mu_1\ldots\mu_n} \psi^i_0 \,
\bar{\psi}^i_0 \Gamma_{(n) ~ \mu_1\ldots\mu_n} \psi^i_0
\label{laggen}
\end{equation}
where $g_{(n)}$ are generalized couplings. Although we will always use
$g$~$=$~$g_{(0)}$ which is not to be confused with the bare coupling constant.
The additional couplings $g_{(n)}$ fall into two classes. For (\ref{lagbare}) 
$g_{(1)}$ and $g_{(2)}$ would correspond to renormalizable interactions in
strictly two dimensions and are not couplings associated with evanescent
interactions. The former coupling corresponds to the Thirring model while the 
latter is related to the interaction $\frac{1}{2} ( \bar{\psi}^i \gamma^5 
\psi^i )^2$ which is part of the chiral Gross-Neveu model. We do not consider 
either of these theories here. Although the projection formalism of 
\cite{26,27} applies equally to their renormalization. The couplings $g_{(n)}$ 
with $n$~$\geq$~$3$ label the set of evanescent operators, defined by    
\begin{equation}
{\cal O}_n ~=~ \half \bar{\psi}^i \Gamma_{(n)}^{\mu_1\ldots\mu_n} \psi^i \,
\bar{\psi}^i \Gamma_{(n) ~ \mu_1\ldots\mu_n} \psi^i ~,
\end{equation}
which are necessary to ensure (\ref{lagbare}) is renormalizable. These
operators will be generated in the renormalization of (\ref{lagbare}) itself.
For (\ref{laggen}) such operators are generated but in the sense that this is 
hidden as their divergences are removed by the renormalization constants 
associated with $g_{(n)}$. For (\ref{lagbare}), \cite{29,30}, the first new
operator, ${\cal O}_{3}$, emerges first at three loops. In other words there 
was a term in the $4$-point function of (\ref{lagbare}) at three loops of the 
form
\begin{equation}
\frac{a_{(3)}}{\epsilon} \Gamma^{\mu\nu\sigma}_{(3)} \otimes 
\Gamma_{(3)\,\mu\nu\sigma}
\label{op3gen}
\end{equation}
where $a_{(3)}$ is the residue of the simple pole in the regularizing parameter
$\epsilon$ with $d$~$=$~$2$~$-$~$2\epsilon$. The tensor product notation is 
understood to mean the different spinor channels into which the operator
$O_{(n)}$ can be decomposed in the $4$-point function. While the operator 
${\cal O}_{(3)}$ is non-existent in strictly two dimensions its generation in 
dimensional regularization cannot be overlooked as its presence will affect the
extraction of the true renormalization group functions. 

There are several ways of considering (\ref{laggen}) from the practical point 
of renormalizing the $SU(N)$ Gross-Neveu model, (\ref{lagbare}). One could take
a general approach beginning with the most general extension of (\ref{lagbare})
which is (\ref{laggen}) with $g_{(1)}$~$=$~$g_{(2)}$~$=$~$0$ at the outset
and then compute all the renormalization constants for the field, mass and the 
coupling constants $g_{(n)}$ with $n$~$\neq$~$1$ and $2$, to four loops. From 
the resultant renormalization group functions then the $\beta$-function for $g$
can be extracted. An alternative approach would be to ignore the full set of 
couplings and instead include the $g$-dependent renormalization constants for 
the various evanescent operators as and when they are generated. The true 
$\beta$-function for $g$ can then be determined using the projection formalism.
The generated operators, such as (\ref{op3gen}), have a hidden effect on the 
renormalization at the first loop order beyond that of when they appear. This 
has to be accounted for in extracting the renormalization group functions which
one would find if the regularization of the two dimensional Lagrangian was not 
in fact dimensional.

To summarize \cite{26,27} there are three essential aspects to the construction
of the renormalization group functions. The first is the determination of what 
are called the naive renormalization group functions which are denoted by 
$\tilde{\gamma}(g)$, $\tilde{\gamma}_m(g)$ and $\tilde{\beta}(g)$ for the wave 
function, mass and coupling constant renormalizations respectively, 
\cite{26,27}. They are constructed in the standard way of renormalizing a 
quantum field theory with the proviso that when an evanescent operator is 
generated, it is included in the Feynman rules for the renormalization at all 
subsequent higher orders. Associated with the evanescent operator generation is
its own $\beta$-function, denoted by $\beta_k(g)$, which is the second aspect 
of the construction where we will use the label $k$ to refer to the strictly 
evanescent quantities and thus $k$~$\geq$~$3$ in two dimensions. For the case 
of (\ref{op3gen}) the residue $a_{(3)}$ is in effect the first term of 
$\beta_3(g)$ which was computed in \cite{29,30,13}. We concentrate here on the 
formalism and give explicit details later. However, in the context of the 
interpretation of (\ref{laggen}) in terms of generated operators without the 
generalized couplings $g_{(k)}$ then to {\em three} loops the dimensionally 
regularized Lagrangian which is used to determine the renormalization group 
functions of the strictly two dimensional theory is, \cite{28,29,30,13}, 
\begin{equation}
L ~=~ i Z_\psi \bar{\psi}^{i} \partialslash \psi^{i} ~-~ m Z_\psi Z_m
\bar{\psi}^{i} \psi^{i} ~+~ \frac{1}{2} g \mu^{2\epsilon} Z_g Z^2_\psi
( \bar{\psi}^i \psi^i )^2 ~+~ \frac{1}{2} g \mu^{2\epsilon} Z_{33} Z^2_\psi
\left( \bar{\psi}^i \Gamma_{(3)}^{\mu\nu\sigma} \psi^i \right)^2 ~.
\label{lag3}
\end{equation}
Here $Z_{33}$ is the counterterm for (\ref{op3gen}) and $\mu$ is the scale
introduced to ensure that the coupling constants are dimensionless in
$d$-dimensions. In (\ref{lag3}) we have used the standard definitions of the
renormalization constants which connect bare with renormalized quantities
which are 
\begin{equation}
\psi_0 ~=~ \psi \sqrt{Z_\psi} ~~~,~~~ m_0 ~=~ m Z_m ~~~,~~~ g_0 ~=~
g Z_g \mu^{2\epsilon} ~.
\label{Zdef}
\end{equation}
In the context of (\ref{laggen}) $Z_{33}$ would correspond to the 
renormalization of the coupling $g_{(3)}$ as would $Z_{nn}$ be the obvious 
generalization for the other couplings $g_{(n)}$. Moreover, all these 
renormalization constants would be functions of the complete set of couplings. 
However, when one considers the reduced non-multiplicatively renormalizable 
Lagrangian (\ref{lagbare}) then only the necessary evanescent operator coupling
constants are included and they will depend only on $g$. This means that the 
first term of $Z_{33}$, for instance, does not begin with unity like $Z_g$.  
 
The final part of the projection formalism is the computation of the underlying
projection functions for the wave function, mass and coupling constant and
denoted by $\rho^{(k)}(g)$, $\rho^{(k)}_m(g)$ and $C^{(k)}(g)$ respectively. 
They are defined and computed through the projection formula, \cite{26,27},
\begin{eqnarray}
\left. \int d^d x \, \NN [ \OO_k ] \right|_{ g_{(i)} = 0 \, , \, d = 2 }
&=& \int d^d x \left( \, \rho^{(k)}(g) \NN [ i
\bar{\psi} \partialslash \psi ~-~ m \bar{\psi}\psi ~+~ 2g \OO_0 ] \right.
\nonumber \\
&& \left. \left. ~~~~~~~~~-~ \rho^{(k)}_m(g) \, \NN[ m \bar{\psi}\psi ] ~+~
C^{(k)}(g) \NN [ \OO_0 ] \right) \right|_{ g_{(i)} = 0 \, , \, d = 2 }
\label{projform}
\end{eqnarray}
where $\NN [ \OO_k ]$ denotes the normal ordering of the evanescent operator 
${\cal O}_k$, $k$~$\geq$~$3$. While we will follow the method of \cite{26,27}
we note that a similar projection method for evanescent $4$-fermi operators was
provided in \cite{34}. In practical terms the relation (\ref{projform}) 
is to be regarded as having meaning only inside a Green's function. For our 
purposes these will be $2$- and $4$-point functions. One evaluates the 
insertion of ${\cal N} [{\cal O}_k]$ in either $n$-point function to a certain 
order in perturbation theory. Then the operators on the right hand side are 
inserted in the same $n$-point function to the same order. After the usual 
operator renormalization the evanescent couplings are set to zero and the 
Green's function determined in strictly two dimensions which is the meaning of 
the restriction ${ g_{(i)} = 0 \, , \, d = 2 }$, \cite{26,27}. The final task 
is to deduce the perturbative expansions of the three projection functions with 
$\rho^{(k)}(g)$ and $\rho^{(k)}_m(g)$ being determined from insertion in a 
$2$-point function and $C^{(k)}(g)$ from the $4$-point function. Finally, once 
the naive renormalization group functions, evanescent operator 
$\beta$-functions and projection functions are known to the loop order
necessary for the order the true two dimensional renormalization group 
functions are needed then these are given by evaluating, \cite{26,27}, 
\begin{eqnarray}
\beta(g) &=& \tilde{\beta}(g) ~+~ \sum_{k=3}^\infty C^{(k)}(g)
\beta_k(g) \nonumber \\
\gamma(g) &=& \tilde{\gamma}(g) ~+~ \sum_{k=3}^\infty \rho^{(k)}(g)
\beta_k(g) \nonumber \\
\gamma_m(g) &=& \tilde{\gamma}_m(g) ~+~ \sum_{k=3}^\infty \rho^{(k)}_m(g)
\beta_k(g) ~.
\label{rgeref}
\end{eqnarray}
In (\ref{rgeref}) the naive renormalization group functions are derived in the
standard fashion through  
\begin{eqnarray}
\tilde{\gamma}(g) &=& \mu \frac{\partial}{\partial \mu} \ln Z_\psi ~~~,~~~
\tilde{\gamma}_m(g) ~=~ -~ \tilde{\beta}(g) \frac{\partial}{\partial g} \ln Z_m
\nonumber \\
\tilde{\beta}(g) &=& (d-2) g ~-~ g \tilde{\beta}(g)
\frac{\partial}{\partial g} \ln Z_g
\end{eqnarray}
once all the renormalization constants have been computed. In 
\cite{28,29,30,13} the three projection functions were evaluated explicitly to 
the requisite loop order to be able to deduce the renormalization group 
functions to four loops. Although in the case of the wave function 
renormalization for (\ref{lagbare}) the leading term of $\rho^{(3)}(g)$ is not 
required since there is no one loop correction to $\gamma(g)$. In the 
determination of $\gamma_m(g)$ at four loops an error was discovered in the
original evaluation of $\beta_3(g)$, \cite{28}, and in a later article 
\cite{35}. Both these articles were three loop analyses and $\beta_3(g)$ had
not been used or tested in the application of the projection formalism to a
four loop computation. While both these articles agreed on the rational part of
$\beta_3(g)$ they differed in the irrational term. This was resolved in 
\cite{13} when $\gamma_m(g)$ was computed to four loops. To achieve this it was
first required to extract the correct value for the irrational part of
$\beta_3(g)$. With the incorrect value of $\beta_3(g)$ the mass anomalous 
dimension does not vanish when $N$~$=$~$\half$ as it ought to since there is no
interaction for this value of $N$. The appearance of this rational value for
$N$ can be understood best if one converts the $SU(N)$ theory, (\ref{lagbare}),
to the case with Majorana fields whence there is an $O(2N)$ symmetry. Thus when
$N$~$=$~$\half$ here the $4$-fermi interaction vanishes due to the Grassmann 
property. Likewise we expect the true $\beta$-function of (\ref{lagbare}) to be
zero when $N$~$=$~$1$. This is because in that case one can use a two 
dimensional Fierz identity to show that 
\begin{equation}
(\bar{\psi} \psi)^2 ~=~ -~ \frac{1}{2} (\bar{\psi} \gamma^\mu \psi)^2 ~.  
\end{equation}
This means that the $N$~$=$~$1$ Lagrangian is equivalent to the abelian
Thirring model whose $\beta$-function is known to be zero, \cite{26,27}. The 
presence of a factor of $(N-1)$ in each term of the three loop $\MSbar$ 
$\beta$-function of (\ref{lagbare}) is already established. However, to this 
order there is no contribution from $\beta_3(g)$ to the true $\beta$-function 
which will first occur at four loops. The emergence of another factor of 
$(N-1)$ at four loops will be an important check on our computations and the
use of the projection formalism. 

\sect{Computational technicalities} 

We devote this section to the technical issues surrounding the evaluation of 
the Feynman diagrams and the organization of the renormalization. In lower loop
renormalization of (\ref{lagbare}) several renormalization group functions were
determined in the massless Lagrangian. For instance, the wave function 
anomalous dimension was derived at three loops in this way, \cite{10}. That was
possible since no infrared problems were introduced in the $2$-point function 
in the massless case and there are no exceptional momentum configurations in
this Green's function. By contrast the computation of the three loop 
$\beta$-function was carried out in several different ways. In \cite{12} the 
interaction of (\ref{lagbare}) was first rewritten in terms of an auxiliary 
field, $\sigma$, before the massive Lagrangian with {\em two} couplings was 
renormalized and the three loop effective potential for $\sigma$ was computed.
Independently in \cite{11} the Lagrangian (\ref{lagbare}) was renormalized 
without introducing an auxiliary field. Later the three loop renormalization 
was revisited in \cite{28,29,30} where the generation of the evanescent 
operator ${\cal O}_3$ was noted. 

While this summarizes the previous higher order loop computations in the 
Gross-Neveu model to proceed to four loops we have followed a more systematic 
algorithm. We will renormalize the $2$- and $4$-point functions of 
(\ref{lagbare}) using the vacuum bubble approach of \cite{36,37} which was 
developed in order to simplify the renormalization of four dimensional 
theories. In this method one expands the massive propagators of (\ref{lagbare})
using the identity
\begin{equation}
\frac{1}{[(k-p)^2+m^2]} ~=~ \frac{1}{[k^2+m^2]} ~+~ 
\frac{[2kp - p^2]}{[k^2+m^2][(k-p)^2+m^2]} ~.
\label{vacid}
\end{equation}
The repeated use of this identity systematically replaces propagators involving
an external momentum, $p$, with propagators involving purely internal loop
momenta $k$. One terminates the expansion using Weinberg's theorem, \cite{38}. 
The algorithm was introduced in \cite{37} for massless gauge theories. However,
applying it to (\ref{lagbare}) with the particle mass already present means 
that an intermediate infrared regularizing mass does not need to be introduced 
in this application unlike \cite{37}. In using this method for (\ref{lagbare}) 
it is virtually a trivial application. It is only in the case of the $2$-point
function when it is multiplied by $\pslash$ and the spinor trace taken that one
has to iterate the identity more than once to the termination point. For the 
extraction of the mass renormalization taking a spinor trace of the $2$-point 
function the first application of (\ref{vacid}) in effect sets the external 
momentum to zero. A similar situation applies in part for the $4$-point 
function where in effect all external momenta are nullified at the outset. As 
part of the vacuum bubble algorithm described in \cite{36,37} the next stage is
the evaluation of the resultant vacuum bubble graphs which emerge. The standard
approach nowadays is to use the Laporta algorithm, \cite{22}. This is a method 
which first constructs relations between Feynman integrals using identities 
established by integration by parts. These algebraic relations can then be 
solved by linear algebra in such a way that all integrals can be expressed in 
terms of a relatively small set of what are called master integrals. The 
$\epsilon$ expansion of these integrals are determined by non-integration by 
parts methods which completes the evaluation of all constituent integrals 
lurking within a Feynman graph of a Green's function. 

We have described the background to the approach of \cite{36,37} as a 
reference for the Gross-Neveu model renormalization here partly as we will make
use of it but mainly because we have had to adapt it given the complication 
with the generation of ${\cal O}_n$. First, what is evident from (\ref{vacid}) 
is that the Feynman integrals have scalar products of the internal and external
momenta in the numerators. As the first part of the Laporta technique these are
replaced by combinations involving the propagators themselves. If there is an 
irreducible scalar product the integral is what is termed completed by the 
inclusion of a propagator not associated with the original topology. This 
additional propagator contains the irreducible scalar product. One feature of 
the Laporta algorithm's power is that it can handle such irreducible 
propagators. While this is the standard procedure there is a complication when 
we consider the $4$-point function of (\ref{lagbare}). If one did not have to 
account for the evanescent operator generation one would take a spinor 
projection of the Green's function to access one channel of the Feynman rule 
and evaluate the resultant Feynman graphs. Instead we have to be more 
systematic and not take any spinor traces for the $4$-point function. This 
means that before we can evaluate any $4$-point function graph we first
disentangle the internal momenta from contractions within $\gamma$-matrix
strings. This leaves Feynman integrals which involve Lorentz tensors of the
internal momenta up to rank $2L$ at $L$ loops. These internal momenta arise 
from the fermionic propagator. However, only an even number of internal momenta
arise since we have already applied the vacuum bubble expansion and a vacuum
bubble integral with an odd number of Lorentz indices is automatically zero.
What remains to be done is to rewrite these Lorentz tensor integrals in terms
of scalar products whence the earlier algorithm can be applied. It transpires
that there is a large number of different combinations of internal momenta in 
the tensor integrals. Rather than develop a result for all possible cases we 
were able to construct a general tensor decomposition up to rank 8 for all
combinations of internal momenta. This is more appropriate given how large a
number occur at four loops. More details on the tensor decomposition is
provided in appendix \ref{app:A}. 

As all the graphs can now be expanded in terms of completely massive scalar
vacuum bubbles the next task is to reduce these to the set of masters. We have 
used the {\sc Reduze} package, \cite{39,40}, which is a C++ coding of the 
Laporta algorithm. It systematically constructs a database of the relations
between all the required integrals and the final masters. One advantage of our 
approach is that we have needed only {\em one} integral family at each loop 
order to cover all possible integrals which arise. To three loops this is 
effectively trivial since the number of propagators in each integral family 
exactly matches the number of independent scalar products of the internal 
momenta. At four loops there are $10$ possible scalar products but since we 
have a single quartic interaction then at most there are eight propagators in a
Feynman graph. So for the four loop integral family we have chosen the ordered 
propagators
\begin{eqnarray}
\label{eq:propags}
&& \left\{ 
\frac{1}{[k_1^2+m^2]},
\frac{1}{[k_2^2+m^2]},
\frac{1}{[k_3^2+m^2]},
\frac{1}{[k_4^2+m^2]},
\frac{1}{[(k_1-k_4)^2+m^2]},
\frac{1}{[(k_2-k_4)^2+m^2]}, ~~~~~~~~~~ \right. \nonumber \\
&& \left. 
\frac{1}{[(k_3-k_4)^2+m^2]},
\frac{1}{[(k_1-k_2)^2+m^2]},
\frac{1}{[(k_1-k_3)^2+m^2]},
\frac{1}{[(k_1-k_2-k_3)^2+m^2]} 
\right\}
\end{eqnarray}  
as the integral family. One of the propagator choices to complete the family
has been chosen to ensure a non-planar topology is covered. The application of
{\sc Reduze} produces a relatively small set of master integrals to be
evaluated and substituted for the evaluation. These are illustrated in Figures
$1$ and $2$. In Figure $2$ each dot on a line represents an increase in the 
power of the original propagator by unity. Although we have only illustrated 
the pure masters in the sense that beyond one loop one can have have products 
of lower loop order masters. For instance, one can have a product of $L$ one 
loop vacuum bubble graphs at each loop order $L$ or at four loops the product 
of the two loop master shown in Figure $1$ emerges as a master in {\sc Reduze}
at four loops. We note that the master integral labelled $1011.1.2$ in Figure
$2$ is the only one which differs from the master basis choice in \cite{41}. 

The final step is the determination of the $\epsilon$ expansion of the 
integrals in Figures $1$ and $2$. Aside from the simple one loop massive vacuum
integral which is trivial to evaluate, at low loop order the leading term of 
the expansion of various integrals has already been found as well as a few 
simple ones at four loops, \cite{13}. However, the known values are not 
sufficient to determine the divergence of the $4$-point function to the simple 
pole in $\epsilon$. This is partly because a subset of the full four loop
masters were needed for the mass anomalous dimension computation and partly due
to spurious poles which emerge when the integration by parts reduction is 
effected. In other words one has sometimes to evaluate a master to several 
finite orders in the $\epsilon$ expansion to ensure that the correct simple 
pole is found for the renormalization constant. Although it transpires that the
higher orders in $\epsilon$ are ordinarily only required for those masters with
a small number of propagators. 

{\begin{figure}[ht]
~~~~~~~~~~~~~
\includegraphics[width=12cm,height=1.7cm]{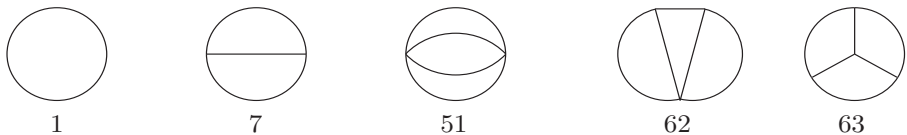}
\caption{One, two and three loop master vacuum bubble integrals.}
\end{figure}}








To find the values of the required terms in the $\epsilon$ expansion of the 
masters we have used the numerical method described in \cite{22,23}. Basically 
master integrals are evaluated numerically to very high precision by solving 
difference equations. The algorithm is not sensitive to the dimension of 
spacetime but needs to be applied separately for each integer. Originally, four 
loop massive vacuum bubbles were numerically determined in four space-time 
dimensions in \cite{42}. 
Subsequently in \cite{41} the analogous three dimensional set of masters was 
also determined to four loops to very high precision for applications to QCD at
finite temperature. In repeating this exercise here at four loops for two 
dimensions, using algorithms developed in \cite{44} as well as \cite{45}, 
we are providing the equivalent machinery which can be applied to 
parallel renormalization calculations in two dimensions. 

However, the ultimate
goal is to have renormalization group functions in terms of rationals as well 
as the Riemann $\zeta$-function. To four loops we do not expect any numbers 
other than those in keeping with our knowledge of the renormalization group 
functions in other theories to the same order in the $\MSbar$ scheme. This is a
key observation mainly because in some of the low loop masters numbers other 
than (Riemann) zeta values appear in the the $\epsilon$ expansion. For 
instance, for the two
loop master in Figure $1$ it is known that special values of the Clausen 
function such as $\mbox{Cl}_2(\pithree)$ contribute (see, for instance, 
\cite{43,13}). In \cite{10,13} the cancellation of this value in the evaluation
of the renormalization constants was checked to three loops. A similar feature 
should emerge here. However, one first has to identify the presence of this and
other such irrationals lurking within the numerical evaluation. This is 
possible through the application of the PSLQ 
algorithm of \cite{24}. Fed with a basis of transcendentals that are expected
to be contained in the numerical value,
the algorithm tries to determine the actual linear combination of those basis 
numbers with rational coefficients. The robustness of the resulting explicit 
relation is tested against a more precise numerical evaluation of the 
coefficient in the $\epsilon$ expansion. With this approach we were able to 
determine all the necessary terms in the $\epsilon$ expansion of the masters of
Figures $1$ and $2$ in order to be able to renormalize the $4$-point function 
of (\ref{lagbare}) to four loops, see appendix \ref{app:B}. 

The re-evaluation of the wave function and mass renormalization constants with 
the masters here was a check on the earlier computations of \cite{10,11,12,13} 
as well as being a partial check on certain values of the 
four loop masters we determined here. It transpired that for the simple pole 
associated with the generation of the new operator ${\cal O}_4$ at four loops a
certain combination of higher order coefficients in the $\epsilon$ expansion of
the masters was required. While each individual coefficient inevitably will 
contain rationals and irrationals aside from $\zeta_n$ we used PSLQ on the 
specific combination which emerged and searched successfully for a linear 
combination using the basis of $\{ 1, \zeta_3, \zeta_4, \zeta_5 \}$. 
This is consistent with our expectations of the basis of numbers which appears 
in an $\MSbar$ renormalization group function at five loops, and the emerging 
linear relation can indeed be used to fix one expansion parameter, see 
(\ref{eq:c30}). The effect the
four loop generation of ${\cal O}_4$ has will become manifest at the next loop 
order similar to the effect ${\cal O}_3$ has in the four loop mass anomalous 
dimension in \cite{13} and the $\beta$-function here. In appendix \ref{app:B} 
we have 
recorded the $\epsilon$ expansion of the masters of Figures $1$ and $2$ to the 
various orders needed for our computations, both numerically and analytically, 
as well as more details of how the master values were found.

{\begin{figure}[ht]
~~~~~~~~~~~~~
\includegraphics[width=12cm,height=8.0cm]{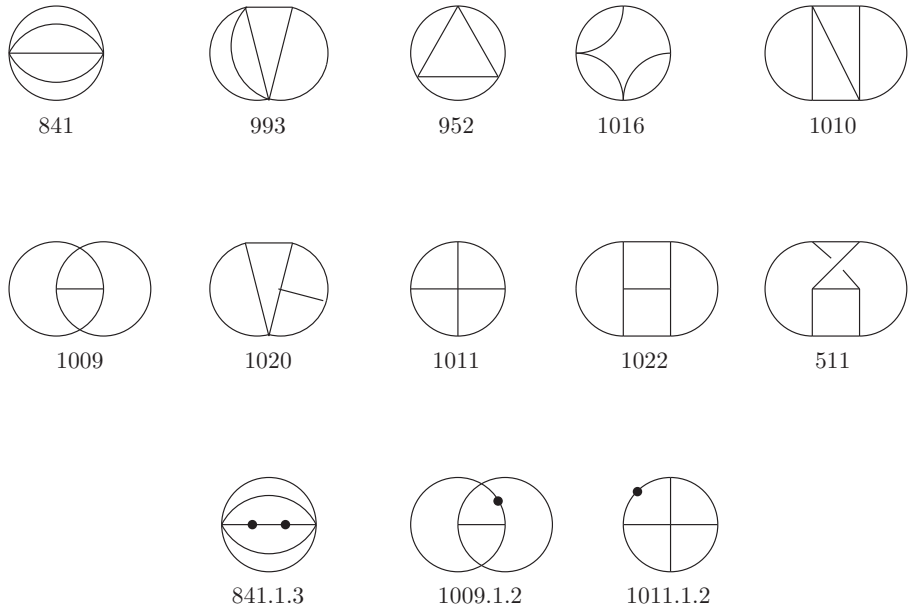}
\caption{Four loop master vacuum bubble integrals. We label them by their 
sector identifier number, whose binary representation corresponds to the 
propagators (out of the set listed in (\ref{eq:propags})) that are present.}
\end{figure}}

Running parallel to the treatment of the masters is the organization of the 
$\gamma$-matrices which had been stripped off to leave tensor integrals. In 
order to extract the naive coupling constant renormalization the 
$\gamma$-matrix strings have to be written in terms of the generalized matrices
$\Gamma_{(n)}^{\mu_1\ldots\mu_n}$. To do this we exploit the algebraic 
properties of these matrices which was discussed at length in \cite{28,29,30}. 
Products of $n$ $\gamma$-matrices can be decomposed into linear combinations of
$\Gamma_{(r)}^{\mu_1\ldots\mu_r}$ where $r$ is even or odd depending on
whether $n$ is even or odd. The range of $r$ begins with $0$ or $1$, depending 
on whether $n$ is even or odd respectively, and ends at $n$. One can construct 
the decomposition iteratively through the basic identities, \cite{28,29,30},
\begin{eqnarray}
\Gamma^{\mu_1 \ldots \mu_n}_{(n)} \gamma^\nu &=&
\Gamma^{\mu_1 \ldots \mu_n \nu}_{(n+1)} ~+~ \sum_{r=1}^n (-1)^{n-r} \,
\eta^{\mu_r \nu} \, \Gamma^{\mu_1 \ldots \mu_{r-1} \mu_{r+1} \ldots
\mu_n}_{(n-1)} \\
\gamma^\nu \Gamma^{\mu_1 \ldots \mu_n}_{(n)} &=&
\Gamma^{\nu \mu_1 \ldots \mu_n}_{(n+1)} ~+~ \sum_{r=1}^n (-1)^{r-1} \,
\eta^{\mu_r \nu} \, \Gamma^{\mu_1 \ldots \mu_{r-1} \mu_{r+1} \ldots
\mu_n}_{(n-1)} ~.
\label{gamid} 
\end{eqnarray}
The process begins with the simple case of $n$~$=$~$1$ which can be written in 
the more familiar way
\begin{equation}
\gamma^\mu \gamma^\nu ~=~ \frac{1}{2} [ \gamma^\mu \gamma^\nu 
- \gamma^\nu \gamma^\mu ] ~+~ \eta^{\mu\nu} I
\end{equation}
where $I$ is the unit matrix in spinor space. This is a simple example of
(\ref{gamid}) and the first term clearly corresponds to
$\Gamma_{(2)}^{\mu\nu}$. With these relations we have constructed the
decomposition of the product of up to $4$ $\gamma$-matrices into the 
generalized $\gamma$-matrices. This is the highest number of $\gamma$-matrices
which can appear in a string at four loops using (\ref{lagbare}) after the 
tensor decomposition of the associated integral has been carried out. There may
be longer strings of $\gamma$-matrices but there will be contractions of at 
least one pair of Lorentz indices within that string. These contractions are 
removed by the systematic application of the Clifford algebra. Then the mapping
of the $\gamma$-matrix string to the matrices $\Gamma_{(n)}^{\mu_1\ldots\mu_n}$
is performed. The totally antisymmetric property of the latter is exploited at 
this stage as the Lorentz index contractions arising from the second terms in
(\ref{gamid}) means that only products of $\gamma$-matrices of the form
$\Gamma^{\mu_1\ldots\mu_n}_{(n)} \otimes \Gamma_{(n)\,\mu_1\ldots\mu_n}$ will
remain. The coefficients of these tensor products will be $d$-dependent due to 
contractions with the $\eta_{\mu\nu}$ tensor and these will therefore impact 
upon the divergence structure of the overall Feynman graph. This is not a 
trivial point. One has to recall that we are dealing with a renormalizable 
quantum field theory which is not multiplicatively renormalizable within 
dimensional regularization. Therefore in the decomposition of the overall 
$\gamma$-algebra to produce the basis of ${\cal O}_n$ operators those labelled 
by $1$ and $2$ cannot emerge with poles in $\epsilon$. These are not evanescent
and ultimately they have to be absent by renormalizability. This is guaranteed 
in effect by factors of $(d-2)$ which emerge as factors of the various 
$n$~$=$~$1$ and $2$ matrices in the $\gamma$-matrix decomposition.

We have now described the technical ingredients of the various aspects of the
computation we have had to perform to renormalize (\ref{lagbare}). From a 
practical point of view the implementation of the procedure and the 
minimization of the significant amounts of algebra could not have been possible
without {\sc Form}, \cite{46,47}. However, at the outset we have used 
{\sc Qgraf}, \cite{48}, to generate all the Feynman graphs for the $2$- and 
$4$-point Green's functions. As we are using a massive version of the 
Lagrangian and applying the vacuum bubble expansion we have to include the 
snail graphs in {\sc Qgraf}. In other words we exclude the {\tt nosnail} option
in the {\tt qgraf.dat} set-up file. Ordinarily we would not highlight such a 
subtlety. However, it is crucial to ensuring the cancellation of the 
non-rational and non-Riemann $\zeta$ numbers in the final renormalization group
functions. The presence of a graph with a snail is to match the mass 
counterterm in a graph at a lower loop order with a similar topology. This was 
observed in earlier work, \cite{10,13}. One consequence of this is that there 
is a larger number of graphs to determine than, say, the equivalent computation 
in scalar $O(N)$ $\phi^4$ theory. There a massless computation can proceed 
without the potential difficulty of infrared problems. In Table $1$ we have 
listed the numbers of graphs for the Green's function we evaluated at each loop
order. Once the graphs have been generated using the {\sc Fortran} coded 
{\sc Qgraf} package, the output is passed to the various {\sc Form} modules 
which add Lorentz, spinor and group indices, apply the Feynman rules and 
prepares the mapping to the integral family notation of {\sc Reduze}. The 
various integrals needed from the {\sc Reduze} database are included in a 
{\sc Form} module before the substitution of the $\epsilon$ expansion of the 
masters and the $d$-dependent factors appearing after the integration by parts.
The final stage is the automatic renormalization of the field and parameters. 
We have followed the method of \cite{49} where the whole computation proceeds
with bare parameters throughout. Only at the end are renormalized quantities
introduced via (\ref{Zdef}) which supplies the necessary counterterms. A caveat
to this procedure is that the renormalization constants for the generated 
evanescent operators have to be included as discussed earlier after the effect 
of ${\cal O}_3$ itself has been allowed for in the one loop $4$-point graphs.  

{\begin{table}[ht]
\begin{center}
\begin{tabular}{|c||c|c|c|c|c|}
\hline
Green's function & $1$ loop & $2$ loop & $3$ loop & $4$ loop & Total \\
\hline
$\psi^i \bar{\psi}^j$ & $1$ & $~2$ & $~~~7$ & $~~~36$ & $~~~46$ \\
$( \bar{\psi}^i \psi^i )^2$
& $3$ & $18$ & $138$ & $1190$ & $1349$ \\
\hline
Total & $4$ & $20$ & $145$ & $1226$ & $1395$ \\
\hline
\end{tabular}
\end{center}
\begin{center}
{Table 1. Number of Feynman diagrams for each $2$- and $4$-point function.}
\end{center}
\end{table}}

\sect{Results}

Having described the technicalities of the computation we now discuss the
results. First, we have reproduced the naive wave function and mass anomalous
dimensions to check with previous work, \cite{1,8,9,10,11,12,13}. We have 
verified that (see section \ref{se:prelim} for definitions)
\begin{eqnarray}
\tilde{\gamma}(g) &=& (2N-1) \frac{g^2}{8\pi^2} ~-~ (N-1)(2N-1)
\frac{g^3}{16\pi^3} ~+~ (4N^2-14N+7)(2N-1) \frac{g^4}{128\pi^4} \nonumber \\
&& +~ \order (g^5) \nonumber \\
\tilde{\gamma}_m(g) &=& -~ (2N-1) \frac{g}{2\pi} ~+~ 
(2N-1) \frac{g^2}{8\pi^2} ~+~ (4N-3)(2N-1) \frac{g^3}{32\pi^3} \nonumber \\
&& +~ \left[ - 40 N^3 - 72 N^2 + 160 N - 81 
+ ( 48 N^3 - 384 N^2 + 492 N - 138 ) \zeta_3 \right] \frac{g^4}{384\pi^4}
\nonumber \\
&& +~ \order (g^5) ~. 
\end{eqnarray}
Equally we have reproduced\footnote{We have corrected the powers of $g$ in our
expressions for $\beta_3(g)$, $\rho^{(3)}_m(g)$ and $C^{(3)}(g)$ which were not
correct in \cite{13,35}.}
\begin{equation}
\rho^{(3)}(g) ~=~ \order (g) ~~~,~~~  
\rho_m^{(3)}(g) ~=~ -~ \frac{1}{\pi} ~+~ \order (g) 
\end{equation}
which together with 
\begin{equation}
\beta_3(g) ~=~ \left[ 3\zeta_3 - 4 \right] 
\frac{g^4}{64\pi^3} ~+~ \order (g^5) 
\end{equation}
which has the corrected irrational term, \cite{13}, means we have rederived 
\begin{eqnarray}
\gamma(g) &=& (2N-1) \frac{g^2}{8\pi^2} ~-~ (N-1)(2N-1)
\frac{g^3}{16\pi^3} ~+~ (4N^2-14N+7)(2N-1) \frac{g^4}{128\pi^4} \nonumber \\
&& +~ \order (g^5) \nonumber \\
\gamma_m(g) &=& -~ (2N-1) \frac{g}{2\pi} ~+~ 
(2N-1) \frac{g^2}{8\pi^2} ~+~ (4N-3)(2N-1) \frac{g^3}{32\pi^3} \nonumber \\
&& +~ ( 2N - 1 ) \left[ - 20 N^2 - 46 N + 57  
+ 12 ( N - 1 ) ( 2N - 13 ) \zeta_3 \right] \frac{g^4}{384\pi^4} ~+~ 
\order (g^5) ~. ~~~~~
\label{rge4}
\end{eqnarray}
It is important to appreciate that these have been derived with the methods 
described in this article using four loop massive vacuum bubbles and the 
Laporta algorithm. In \cite{50} the four loop wave function was evaluated in 
the completely massless theory, and verified in \cite{51}, while the mass 
anomalous dimension was derived with massive vacuum bubbles but did not use the
Laporta algorithm. Instead the necessary terms of the $\epsilon$ expansion of 
the various masters, which were fewer than those required for the 
$\beta$-function at four loops, could be derived from known four dimensional 
ones and related to those in two dimensions using Tarasov's method, 
\cite{52,53}. For completeness we have computed the next term in the series for
$\beta_3(g)$ and found  
\begin{equation}
\beta_3(g) ~=~ \left[ 3\zeta_3 - 4 \right] 
\frac{g^4}{64\pi^3} ~+~ \left[ - 24 N \zeta_3 - 18 N \zeta_4 + 56 N - 24 \zeta_3
+ 9 \zeta_4 + 2 \right] \frac{g^5}{384\pi^4} ~+~ \order (g^6) ~. 
\end{equation}
Equally we find that 
\begin{equation}
\beta_4(g) ~=~ \left[ 108 \zeta_3 + 18 \zeta_4 - 30 \zeta_5 - 107 \right] 
\frac{g^5}{1536\pi^4} ~+~ \order (g^6)
\label{eq:betaop4}
\end{equation}
for the generated ${\cal O}_4$ operator at four loop. The $\order (g^5)$ terms 
of each of these $\beta$-functions will be required for the five loop mass and
$\beta$-function computations and are consistent with expectations that
$\zeta_4$ can only appear first at five loops in the renormalization group
functions.

For the similar derivation of the true $\beta$-function we note that the naive 
version which we have calculated here is 
\begin{eqnarray}
\tilde{\beta}(g) &=& (d-2)g ~-~ ( N - 1 ) \frac{g^2}{\pi} ~+~
( N - 1 ) \frac{g^3}{2\pi^2} ~+~ ( N - 1 ) ( 2N - 7 ) \frac{g^4}{16\pi^3} 
\nonumber \\
&& +~ \left[ - 4 N^3 - 34 N^2 + 86 N - 60 - ( 132 N^2 - 336 N + 195 ) \zeta_3
\right] \frac{g^5}{96\pi^4} ~+~ \order (g^6) ~~~
\end{eqnarray}
where it is clear that there is no $(N-1)$ factor at four loops. Though it is
worth noting that at $N$~$=$~$1$ the four loop coefficient is proportional to 
$[ 4 - 3 \zeta_3] \frac{1}{32\pi^4}$. The emergence of this combination of
rational and irrational numbers which is proportional to the coefficient of the
leading term of $\beta_3(g)$ is indicative that the naive $\beta$-function is
not incorrect. With the evaluation of 
\begin{equation}
C^{(3)}(g) ~=~ -~ \frac{2}{\pi} g ~+~ \order (g^2) 
\end{equation}
we finally find our main result that 
\begin{eqnarray}
\beta(g) &=& (d-2)g ~-~ ( N - 1 ) \frac{g^2}{\pi} ~+~
( N - 1 ) \frac{g^3}{2\pi^2} ~+~ ( N - 1 ) ( 2N - 7 ) \frac{g^4}{16\pi^3} 
\nonumber \\
&& +~ ( N - 1 ) \left[ - 2 N^2 - 19 N + 24 - 6 \zeta_3 ( 11 N - 17 )
\right] \frac{g^5}{48\pi^4} ~+~ \order (g^6)
\label{beta4}
\end{eqnarray}
to four loops in the $\MSbar$ scheme. 

Aside from the correct appearance of $(N-1)$ in each term there are several
other independent checks. One of these is the correct value of the residue of
the poles in $\epsilon$ beyond the simple one in $Z_g$. These non-simple pole
residues are already determined from the values of the residues of all the 
poles of the lower loop order parts of $Z_g$ from the renormalization group
equation. The other main check is through the large $N$ expansion. This is
an alternative but complementary way of determining the coefficients in the
perturbative series of a renormalization group function. Briefly the method of 
computing large $N$ information relies on the renormalization group equation 
considered at the Wilson-Fisher fixed point in $d$-dimensions. In 
\cite{54,55,56} a method was developed to evaluate the critical exponents in
the fixed point universal theory in the $1/N$ expansion where $N$ is large. The
information encoded in the various critical exponents can be extracted and have
a direct relation to the coefficients of the polynomials in $N$ of the 
corresponding renormalization group function at each loop order. In other words
before we had computed (\ref{beta4}) from the application of the method of
\cite{54,55,56} to the case of (\ref{lagbare}), \cite{57,58,59,60,61,62,63}, 
the coefficients of the cubic and quadratic terms in $N$ had already been 
predicted. We made no assumption at the outset as to what these values would be 
in this computation. Their emergence from the full evaluation consistent with 
the critical exponent corresponding to the $\beta$-function slope at $\order 
(1/N^2)$ is a non-trivial check on (\ref{beta4}).  

\sect{Applications}

We now turn to several applications of the results and examine the critical 
exponents derived from the renormalization group functions in various cases as 
we can now deduce the Wilson-Fisher fixed point location at four loops. The 
first situation we consider is when $N$~$=$~$4$ which lies in the chiral Ising 
universality class, \cite{64}, and is related to a particular electronic phase 
transition in the honeycomb lattice of graphene. Indeed it is noted in 
\cite{64} that this transition from semi-metal to a Mott insulator could be a 
mimic of spontaneous symmetry breaking in the Standard Model and moreover can 
be studied in principle in the laboratory. In \cite{64} estimates for the 
exponents $\eta_\psi$ and $\nu$ as well as that denoted by $\eta_\phi$ were 
given in three dimensions using functional renormalization group methods, by 
summing the $\epsilon$ expansion of the Gross-Neveu model above two dimensions,
the Gross-Neveu-Yukawa model below four dimensions as well as the three 
dimensional large $N$ exponents, \cite{64}. As only three loop renormalization 
group functions were available for the Gross-Neveu case it is appropriate to 
extend that analysis here. 
The critical exponents are defined in terms of our renormalization group 
functions at criticality, \cite{64}, by 
\begin{equation}
\eta_\psi ~=~ \gamma(g_c) ~~~,~~~ \eta_\phi ~=~ d ~+~ 2 \gamma_m(g_c) ~~~,~~~
\frac{1}{\nu} ~=~ -~ \beta^\prime(g_c) ~. 
\end{equation}
In order to make the comparison with the notation of
\cite{64} easier we set $d$~$=$~$2$~$+$~$\varepsilon$ for the moment and find
\begin{eqnarray}
\eta_\psi &=& \frac{7}{72} \varepsilon^2 ~-~ \frac{7}{432} \varepsilon^3 ~+~
\frac{7}{10368} \varepsilon^4 ~+~ \order (\varepsilon^5) \nonumber \\
\eta_\phi &=& 2 ~-~ \frac{4}{3} \varepsilon ~-~ \frac{7}{36} \varepsilon^2 ~+~
\frac{7}{54} \varepsilon^3 ~+~ 
\frac{91 [12 \zeta_3 + 1 ]}{5184} \varepsilon^4 ~+~ 
\order (\varepsilon^5) \nonumber \\
\frac{1}{\nu} &=& \varepsilon ~-~ \frac{1}{6} \varepsilon^2 ~-~ 
\frac{5}{72} \varepsilon^3 ~+~ \frac{[81 \zeta_3 + 35]}{216} \varepsilon^4 ~+~ 
\order (\varepsilon^5) 
\end{eqnarray} 
where
\begin{equation}
\frac{g_c}{\pi} ~=~ \frac{1}{3} \varepsilon ~+~ \frac{1}{18} \varepsilon^2 ~+~
\frac{1}{48} \varepsilon^3 ~+~ 
[ - 108 \zeta_3 - 31 ] \frac{\varepsilon^4}{2592} ~+~
\order (\varepsilon^5) ~.
\end{equation} 
We note that we have evaluated our renormalization group functions with 
$N$~$=$~$4$ rather than $N$~$=$~$2$ as in \cite{64} since we have used
$\mbox{Tr} I$~$=$~$2$ for our $\gamma$-matrix trace normalization in contrast
to the rank $4$ $\gamma$-matrices used in \cite{64}. The spinor trace always
appears within graphs with a closed fermion loop which has a factor of $N$
deriving from the $SU(N)$ symmetry. Numerically we have 
\begin{eqnarray}
\eta_\psi &=& 0.097222 \varepsilon^2 ~-~ 0.016204 \varepsilon^3 ~+~ 
0.000675 \varepsilon^4 ~+~ \order (\varepsilon^5) \nonumber \\
\eta_\phi &=& 2 ~-~ 1.333333 \varepsilon ~-~ 0.194444 \varepsilon^2 ~+~ 
0.129630 \varepsilon^3 ~+~ 0.270765 \varepsilon^4 ~+~ \order (\varepsilon^5) 
\nonumber \\
\frac{1}{\nu} &=& \varepsilon ~-~ 0.166667 \varepsilon^2 ~-~ 
0.069444 \varepsilon^3 ~+~ 0.612808 \varepsilon^4 ~+~ \order (\varepsilon^5) ~.
\label{numexp4}
\end{eqnarray}
Aside from $\eta_\psi$ the four loop corrections are larger (at $\varepsilon=1$) than the three loop
ones. To gain estimates of these exponents in three dimensions we have used 
Pad\'{e} approximants and noted the results in Table 2. There we used the 
leading value for the two loop estimate of $\eta_\psi$ since there is no one 
loop term. The estimates for $\eta_\phi$ are from an $[0,L]$ approximant at $L$
loops and those for $1/\nu$ are deduced from summing $\epsilon \nu$ and then 
inverting. As the four loop correction to $\eta_\psi$ is virtually zero the 
estimate is stable. However, the value is twice that of the functional 
renormalization group analysis of \cite{64}. For the other two exponents the 
three loop estimates differ from those quoted in Table I of \cite{64}. This is 
because as far as we can tell those values seem to be determined by setting 
$\varepsilon$~$=$~$1$ explicitly in the expansion rather than using an
approximant as was the case for the large $N$ estimates, \cite{64}. In our
situation the three loop estimate for $\eta_\phi$ is in keeping with the large 
$N$ and functional renormalization group values of $0.760$-$0.776$. Although 
our four loop value is closer to the Monte-Carlo simulation of $0.745(8)$ of 
\cite{65} it appears that convergence has not been reached unlike $\eta_\psi$. 
There seems to be a similar situation for $1/\nu$ since our estimates are 
oscillating although the four loop value is competitive with the range 
$0.949$-$0.995$ given in Table I of \cite{64} and the Monte-Carlo estimate of 
$1.00(4)$ of \cite{65}. A more comprehensive fixed point analysis of extended 
Gross-Neveu type models has been provided recently for spinless fermions on a 
honeycomb lattice in \cite{66}.

{\begin{table}[ht]
\begin{center}
\begin{tabular}{|c||c|c|c|c|}
\hline
Exponent & $2$ loop & $3$ loop & $4$ loop & MC estimate\\
\hline
$\eta_\psi$ & $0.097$ & $0.083$ & $0.082$ & --\\ 
$\eta_\phi$ & $0.906$ & $0.778$ & $0.745$ & $0.745(8)$~\cite{65}\\ 
$1/\nu$ & $0.857$ & $0.784$ & $0.931$ & $1.00(4)$~\cite{65}\\ 
\hline
\end{tabular}
\end{center}
\begin{center}
{Table 2. Estimates for critical exponents when $N$~$=$~$4$ using Pad\'{e}
approximants, compared with Monte-Carlo results from the literature.}
\end{center}
\end{table}}

A second application is to the case of the replica limit $N$~$\to$~$0$ which
has been examined in \cite{67,68,69} for other graphene related problems but is
also relevant to the random bond Ising model problem, \cite{4}. For example, 
the three dimensional theory in the replica limit describes the transition from
a relativistic semi-metal to a diffusive metallic phase. From (\ref{rge4}) and 
(\ref{beta4}) we find that  
\begin{eqnarray}
\left. \beta(g) \right|_{N=0} &=& (d-2)g 
~+~ \frac{g^2}{\pi} ~-~ \frac{g^3}{2\pi^2} ~+~ \frac{7g^4}{16\pi^3} ~+~ 
\frac{[ - 17 \zeta_3 - 4]}{8\pi^4} g^5 ~+~ \order (g^6) \nonumber \\
\left. \gamma(g) \right|_{N=0} &=& -~ \frac{g^2}{8\pi^2} ~-~ 
\frac{g^3}{16\pi^3} ~-~ \frac{7g^4}{128\pi^4} ~+~ \order (g^5) \nonumber \\
\left. \gamma_m(g) \right|_{N=0} &=& 
\frac{g}{2\pi} ~-~ \frac{g^2}{8\pi^2} ~+~ \frac{3g^3}{32\pi^3} ~+~ 
\frac{[- 52 \zeta_3 - 19 ]}{128\pi^4} g^4 ~+~ \order (g^5) ~.
\end{eqnarray}
Solving for $g_c$ from $\beta(g_c)$~$=$~$0$, and reverting to 
$d$~$=$~$2$~$-$~$2\epsilon$ again, leads to
\begin{eqnarray}
\left. \gamma(g_c) \right|_{N=0} &=& -~ \frac{1}{2} \epsilon^2 ~-~
\frac{3}{2} \epsilon^3 ~-~ \frac{25}{8} \epsilon^4 ~+~ \order (\epsilon^5) 
\nonumber \\
\left. \gamma_m(g_c) \right|_{N=0} &=& \epsilon ~+~ \frac{1}{2} \epsilon^2 ~+~ 
\frac{1}{8} [ 84 \zeta_3 - 5 ] \epsilon^4 ~+~ \order (\epsilon^5) \nonumber \\
\left. \beta^\prime(g_c) \right|_{N=0} &=& 2 \epsilon ~-~ 2 \epsilon^2 ~+~ 
3 \epsilon^3 ~-~ 6 [ 17 \zeta_3 + 1 ] \epsilon^4 ~+~ \order (\epsilon^5) 
\label{repexp}
\end{eqnarray}
or, in numerical form,
\begin{eqnarray}
\left. \gamma(g_c) \right|_{N=0} &=& -~ 0.500000 \epsilon^2 ~-~ 
1.500000 \epsilon^3 ~-~ 3.125000 \epsilon^4 ~+~ \order (\epsilon^5)
\nonumber \\
\left. \gamma_m(g_c) \right|_{N=0} &=& \epsilon ~+~ 0.500000 \epsilon^2 ~+~ 
11.996597 \epsilon^4 ~+~ \order (\epsilon^5) \nonumber \\
\left. \beta^\prime(g_c) \right|_{N=0} &=& 2.000000 \epsilon ~-~ 
2.000000 \epsilon^2 ~+~ 3.000000 \epsilon^3 ~-~ 128.609804 \epsilon^4 ~+~ 
\order (\epsilon^5) ~.~~
\end{eqnarray}
While there is no $\order (\epsilon)$ term for the wave function exponent since
$\gamma(g)$ begins at one loop it is unusual that there is no $\order 
(\epsilon^3)$ term in $\gamma_m(g_c)$. 

In principle we could repeat our $N$~$=$~$4$ exercise of summing each series.
However from (\ref{repexp}) it turns out that each four loop coefficient is 
rather large in comparison with the three loop term, unlike the previous
application, and indicates that even setting $\epsilon$~$=$~$-$~$\half$ will
lead to diverging series. We have tried various resummation methods, such as 
Pad\'{e} approximants, in order to improve convergence but have not found any 
credible exponent estimates. While this appears to be disappointing and 
different from the situation where summing $\epsilon$ expansions in other 
theories, or $N$~$=$~$4$ here, has led to reliable exponent estimates it is in
accord with recent observations in \cite{69}. In \cite{69} it was noted that in 
trying to apply the perturbative $\epsilon$ expansion results to three 
dimensions there may be contributions from some or all of the evanescent
operators ${\cal O}_n$ to the value of the exponents when 
$\epsilon$~$=$~$-$~$\half$. For instance, near two dimensions omitting the 
contribution from ${\cal O}_3$ would have led to an inconsistency with the 
symmetry of the theory. However, in three dimensions this operator is not 
excluded and is not unrelated to the pseudo-tensor $\epsilon^{\mu\nu\sigma}$. 
By contrast ${\cal O}_4$ would remain evanescent. A possible resolution for the
application of the $\epsilon$ expansion to the three dimensional replica limit 
problem would be to consider another theory in the same universality class to 
obtain reliable estimates such as the four dimensional Gross-Neveu-Yukawa 
theory, \cite{67,68,69}. An alternative would be to extend our analysis here to
renormalize (\ref{laggen}) and determine the $\beta$-functions of all the 
couplings up to a certain loop order. From these the fixed point structure 
could be analysed to see if ${\cal O}_3$, for example, influenced the 
convergence of the $\epsilon$ expansion of the exponents in the approach to 
three dimensions. That is clearly beyond the scope of the present article. 
However, if there is an evanescent operator issue in estimating exponents in 
the $\epsilon$ expansion it is not immediately apparent in the $N$~$=$~$4$ 
case. Although the four loop corrections in (\ref{lagbare}) are larger then the
three loop ones we were able to find estimates in reasonable agreement with 
other methods. For $N$~$=$~$4$ any breakdown may not become apparent until 
$\order (\epsilon^5)$. 
 
\sect{Discussion}

We conclude with brief remarks. The article represents the completion of the
four loop renormalization of the two-dimensional $SU(N)$ Gross-Neveu model. It 
has been a technical exercise due to the need to handle the problem of the 
generation of evanescent $4$-fermi operators in the dimensional regularization 
of the Lagrangian. What has been reassuring is that the formalism of 
\cite{26,27} has been robust and we were able to extract the $\beta$-function 
as our main result in (\ref{beta4}), consistent with the vanishing of the 
$\beta$-function for the abelian Thirring model. Not properly taking into 
account the effect of the evanescent operators would have led to an 
inconsistent result.

We have also provided the $\beta$-function for 
the new evanescent operator ${\cal O}_4$ which will be necessary for any future
five loop renormalization. Such a computation is now viable in principle in 
part since the approach provided here can clearly be extended to the next
order. Also because the calculational technology, such as the Laporta algorithm
to reduce Feynman integrals by integration by parts and then evaluate masters 
numerically, is available. Indeed there has been substantial recent progress in
evaluating five loop massive vacuum bubbles for renormalization group functions
of QCD, \cite{45,70}. Although there are potential obstacles such as the actual
computer programming and running times these are not insurmountable issues as 
is evident from recent progress in $\beta$-function evaluation in similar 
models mentioned earlier. 

Another issue which would be intriguing to
investigate is the effect evanescent operators have on the $\epsilon$ 
expansion of exponents. Given the interest in the connection of AdS/CFT with
simple $O(N)$ scalar and Gross-Neveu models, \cite{7}, understanding how such 
extra operators are manifest in the equivalence of theories in the same 
universality class at the Wilson-Fisher point in $d$-dimensions may prove 
useful in conformal bootstrap studies such as that of \cite{6}. Such a study 
would not be an isolated investigation. For instance, given the functional 
renormalization group analyses of more general Gross-Neveu models, \cite{66}, 
having the four loop perturbative renormalization group functions for that 
generalization would be useful for refining the fixed point structure. 

\vspace{1cm}
\noindent
{\bf Acknowledgements.} 
This work was supported in part by the STFC Consolidated Grant number 
ST/L000431/1, DFG grant SCHR 993/2, FONDECYT project 1151281 and UBB 
project GI-152609/VC. 
One of the authors (JAG) also thanks the Galileo 
Galilei Institute for Theoretical Physics for hospitality and the INFN for 
partial support during the completion of part of this work. All diagrams were 
drawn with Axodraw, \cite{71}.

\appendix 

\sect{Tensor reduction} 
\label{app:A}

We provide background concerning the tensor reduction required for the massive 
vacuum integrals in this appendix. Lower rank examples have already been given 
in \cite{13} and we made use of these here. For instance, 
\begin{eqnarray}
&& \int_k k^{\mu_1} k^{\mu_2} f_1(k^2) ~=~ \frac{\eta^{\mu_1\mu_2}}{d} \int_k 
k^2 f_1(k^2) \nonumber \\
&& \int_{kl} k_1^{\mu_1} k_2^{\mu_2} k_3^{\mu_3} k_4^{\mu_4} f_2(k,l)
\nonumber \\
&& =~ \frac{1}{d(d-1)(d+2)} 
\int_{kl} \left\{ \!\! \frac{}{}
\left[ (d+1) k_1.k_2 k_3.k_4 - k_1.k_3 k_2.k_4
- k_1.k_4 k_2.k_3 \right]
\eta^{\mu_1\mu_2} \eta^{\mu_3\mu_4} \right. \nonumber \\
&& \left. ~~~~~~~~~~~~~~~~~~~~~~~~~~~~~~+  
\left[ (d+1) k_1.k_3 k_2.k_4 - k_1.k_2 k_3.k_4
- k_1.k_4 k_2.k_3 \right]
\eta^{\mu_1\mu_3} \eta^{\mu_2\mu_4} \right. \nonumber \\
&& \left. ~~~~~~~~~~~~~~~~~~~~~~~~~~~~~~+  
\left[ (d+1) k_1.k_4 k_2.k_3 - k_1.k_2 k_3.k_4
- k_1.k_3 k_2.k_4 \right]
\eta^{\mu_1\mu_4} \eta^{\mu_2\mu_3} \frac{}{} \right\} f_2(k,l) 
\nonumber \\ 
&& \int_{klq} k_1^{\mu_1} k_2^{\mu_2} k_3^{\mu_3} k_4^{\mu_4} k_5^{\mu_5} 
k_6^{\mu_6} f_3(k,l,q) \nonumber \\ 
&& =~ \frac{\eta^{\mu_1\mu_2} \eta^{\mu_3\mu_4} \eta^{\mu_5\mu_6}}
         {d(d-1)(d-2)(d+2)(d+4)} \,\times\nonumber \\
&& ~~~ \times \int_{klq} \!\! \left\{ \frac{}{} 
(d^2+3 d-2) k_1.k_2 k_3.k_4 k_5.k_6   
           - (d+2) k_1.k_2 k_3.k_5 k_6.k_4  
            - (d+2) k_1.k_2 k_3.k_6 k_4.k_5   
            \right. \nonumber \\
&& \left. ~~~~~~~~~~~~~ 
            - (d+2) k_1.k_3 k_2.k_4 k_5.k_6  
            + 2 k_1.k_3 k_2.k_5 k_6.k_4   
            + 2 k_1.k_3 k_2.k_6 k_4.k_5 
            \right. \nonumber \\
&& \left. ~~~~~~~~~~~~~ 
            - (d+2) k_1.k_4 k_2.k_3 k_5.k_6   
            + 2 k_1.k_4 k_2.k_5 k_6.k_3  
            + 2 k_1.k_4 k_2.k_6 k_3.k_5  
            \right. \nonumber \\
&& \left. ~~~~~~~~~~~~~ 
            + 2 k_1.k_5 k_2.k_3 k_4.k_6  
            + 2 k_1.k_5 k_2.k_4 k_6.k_3   
            - (d+2) k_1.k_5 k_2.k_6 k_3.k_4 
            \right. \nonumber \\
&& \left. ~~~~~~~~~~~~~ 
            + 2 k_1.k_6 k_2.k_3 k_4.k_5   
            + 2 k_1.k_6 k_2.k_4 k_5.k_3  
            - (d+2) k_1.k_6 k_2.k_5 k_3.k_4 \frac{}{} \right\} f_3(k,l,q) 
            \nonumber \\
&& ~~~+~ \mbox{$14$ similar terms} 
\label{decomp}
\end{eqnarray} 
and we refrain from listing the rank 8 reduction.
In each decomposition the internal momenta $k_i$ are in the set of loop momenta
defined by the subscript on the integral symbol and
\begin{equation}
\int_k ~=~ \int \frac{d^dk}{(2\pi)^d} ~. 
\end{equation} 
In this appendix we have indicated the scalar products between vectors by
including the dot. The functions $f_i(k_1,\ldots,k_i)$ formally represent any 
integrand involving massive propagators of the internal momenta. One feature
which emerges in the rank $6$ decomposition is the appearance of a spurious
pole in two dimensions. This is different from similar poles which emerge in
the integration by parts but has the same consequence which is that we require
the master integrals to higher order in $\epsilon$ than would naively be
expected. There is also a factor of $1/(d-2)$ in the rank $8$ decomposition.
However as the expression for the decomposition which is used at three 
loops is quite large we do not provide the explicit form which was required at 
four loops. That involves $105$ different combinations of four products of the 
$\eta^{\mu\nu}$ tensor. This and the lower rank decompositions were constructed
by a projection method. The tensor integral is written as a linear combination
of all possible products of the metric tensor. Then the coefficients are 
determined by inverting the matrix derived by systematically multiplying the
integral by one of the rank $2L$ product of metric tensors. At four loops this
is a $105$~$\times$~$105$ matrix and each entry is a power of $d$. Ultimately
the coefficients in the decomposition involve scalar products of the loop
momenta as is evident in (\ref{decomp}). One benefit in using the most general 
decomposition at each even rank is that the lower rank ones are still 
applicable at higher loop order but more importantly it simplifies the amount 
of effort required in the computation itself. In coding these relations in 
{\sc Form}, \cite{46}, and {\sc Tform}, \cite{47}, we exploited the set 
facility of that language. In other words the actual internal momenta $k$, $l$ 
and $q$ for instance are contained within a larger set of declared momentum 
vectors which also include the wildcard internal momenta $k_i$. With one 
{\sc Form} identification at each even rank then {\em all} possible 
combinations of tensor integrals which can arise can be substituted 
immediately. The outcome is that the integrals are now all in the scalar 
product form to which the Laporta algorithm can then be applied. It is worth 
recalling that the presence of a mass at the outset ensures that we are in an 
infrared safe scenario in the extraction of the poles in $\epsilon$. 

\sect{Master integrals} 
\label{app:B}

In this appendix we record the explicit values of the various
$d$~$=$~$2$~$-$~$2\epsilon$ dimensional 
massive master vacuum bubble integrals to four loops. These expressions are
parallel to similar $\epsilon$ expansion of master integrals in three, 
\cite{41}, and four, \cite{42}, dimensions. Although it is worth noting that 
the master basis in those papers is not the same in each dimension. The first
stage in determining the integrals we require is their $\epsilon$ expansion in
high precision numerical form, which we obtain using algorithms developed in 
\cite{44}. Using the notation $I_i$ for each integral where
$i$ corresponds to the label defining the graphs in Figures $1$ and $2$ then we
have 
\begin{eqnarray}
I_{7} &=& 
 +~ 2.3439072386894588906015622888722770690954030096358\,\epsilon^{2}
 \nn\\&&-~ 4.0375761317658220051348256207336882506736789108211\,\epsilon^{3}
 \nn\\&&+~ 8.0425620153544679772236993285907591563591747215790\,\epsilon^{4}
 \nn\\&&-~ 16.025786102703578704072144195119000589039259586708\,\epsilon^{5}
 \nn\\&&+~ 32.016942887378604993187557907045446315854486134669\,\epsilon^{6}
 \nn\\&&+~ \order(\!\,\epsilon^{7})
\\ I_{51} &=& 
 +~ 8.4143983221171599977981671305801499353549040463834\,\epsilon^{3}
 \nn\\&&-~ 39.381582123122165624939809345086245359050066808606\,\epsilon^{4}
 \nn\\&&+~ 147.66098871791534853160710390728010530078749831966\,\epsilon^{5}
 \nn\\&&-~ 503.10764699842011083276305336842813338479312345530\,\epsilon^{6}
 \nn\\&&+~ \order(\!\,\epsilon^{7})
\\ I_{62} &=& 
 +~ 1.4691806594172819043774577529584388776016499128606\,\epsilon^{3}
 \nn\\&&-~ 1.3063884197794647578841243370437552845041356582719\,\epsilon^{4}
 \nn\\&&+~ 2.0331116532991875366360222945013840018039180084787\,\epsilon^{5}
 \nn\\&&-~ 2.9739730075270498775921859477377291738249043348203\,\epsilon^{6}
 \nn\\&&+~ \order(\!\,\epsilon^{7})
\\ I_{63} &=& 
 +~ 0.4006856343865314284665793871704833302549954307801\,\epsilon^{3}
 \nn\\&&+~ 0.4835885331655307913305885127498895695285786668497\,\epsilon^{4}
 \nn\\&&+~ 0.0780252050755395162371825361501387815082130272829\,\epsilon^{5}
 \nn\\&&+~ 0.0375891517515336496720441119688792207019879946984\,\epsilon^{6}
 \nn\\&&+~ \order(\!\,\epsilon^{7})
\\ I_{841} &=& 
 +~ 39.945588011996245698944028498423995891849782924769\,\epsilon^{4}
 \nn\\&&-~ 348.30111814405516255151518000519957122301660173275\,\epsilon^{5}
 \nn\\&&+~ 2109.3709165841690773369924824477570911886774067257\,\epsilon^{6}
 \nn\\&&+~ \order(\!\,\epsilon^{7})
\\ I_{993} &=& 
 +~ 4.3469994124634535378088800462728468761816547339772\,\epsilon^{4}
 \nn\\&&-~ 13.577707070928288751118628100406503145199518907347\,\epsilon^{5}
 \nn\\&&+~ 37.127160597831325629352443903530938622368671294815\,\epsilon^{6}
 \nn\\&&+~ \order(\!\,\epsilon^{7})
\\ I_{952} &=& 
 +~ 3.6061707094787828561992144845343499722949588770214\,\epsilon^{4}
 \nn\\&&-~ 9.8138144306011662120436866789897530527489837499043\,\epsilon^{5}
 \nn\\&&+~ 24.411671104819532543287943944108057148529562577004\,\epsilon^{6}
 \nn\\&&+~ \order(\!\,\epsilon^{7})
\\ I_{1016} &=& 
 +~ 1.1103912916056864445568224113958020118863386462471\,\epsilon^{4}
 \nn\\&&-~ 0.5570931353458396260454773430999778874389836524579\,\epsilon^{5}
 \nn\\&&+~ 0.9538577567535969561534040876837052542479617978881\,\epsilon^{6}
 \nn\\&&+~ \order(\!\,\epsilon^{7})
\\ I_{1010} &=& 
 +~ 1.0189569061909011214495005968296777382762004059847\,\epsilon^{4}
 \nn\\&&-~ 0.3771202916485063542254503885080026978927612968436\,\epsilon^{5}
 \nn\\&&+~ 0.7375177343111526021741885922982090183281771921329\,\epsilon^{6}
 \nn\\&&+~ \order(\!\,\epsilon^{7})
\\ I_{1009} &=& 
 +~ 0.8565905156509505479411015870020216902045909958594\,\epsilon^{4}
 \nn\\&&-~ 0.1213313173697959016400906654647564495669736138643\,\epsilon^{5}
 \nn\\&&+~ 0.4842932258265766625264348872290029324911909578604\,\epsilon^{6}
 \nn\\&&+~ \order(\!\,\epsilon^{7})
\\ I_{1020} &=& 
 +~ 0.3208023444820586600649990118869225971586777108797\,\epsilon^{4}
 \nn\\&&+~ 0.4572462611421325605718900723432215797880749696720\,\epsilon^{5}
 \nn\\&&+~ 0.1586006061868626137523017429084224535466048502067\,\epsilon^{6}
 \nn\\&&+~ \order(\!\,\epsilon^{7})
\\ I_{1011} &=& 
 +~ 0.2628085426500804665114416461835485085723719210284\,\epsilon^{4}
 \nn\\&&+~ 0.4576418734183239967463989327187398356783555081320\,\epsilon^{5}
 \nn\\&&+~ 0.2029039559648421469240439142252097171397779702214\,\epsilon^{6}
 \nn\\&&+~ \order(\!\,\epsilon^{7})
\\ I_{1022} &=& 
 +~ 0.1117150037641884515592837802131386089989996846176\,\epsilon^{4}
 \nn\\&&+~ 0.3310065843046395535001459002764848158576544106555\,\epsilon^{5}
 \nn\\&&+~ 0.3408959619840503762721134319337008020593896711925\,\epsilon^{6}
 \nn\\&&+~ \order(\!\,\epsilon^{7})
\\ I_{511} &=& 
 +~ 0.1009095634750077403668961034423758951141639097300\,\epsilon^{4}
 \nn\\&&+~ 0.3117547219541633763351461670518867617161781823227\,\epsilon^{5}
 \nn\\&&+~ 0.3428043862436104836764054879020396593614543308941\,\epsilon^{6}
 \nn\\&&+~ \order(\!\,\epsilon^{7})
\\ I_{841.1.3} &=& 
 +~ 4.1419068549087621355895544541703994542221894809501\,\epsilon^{4}
 \nn\\&&-~ 15.381626198423314746507150138351697129785646284184\,\epsilon^{5}
 \nn\\&&+~ 54.042164666976954209837796635081440969706812114375\,\epsilon^{6}
 \nn\\&&+~ \order(\!\,\epsilon^{7})
\\ I_{1009.1.2} &=& 
 +~ 0.3837923103720091360393035310183704585295526060290\,\epsilon^{4}
 \nn\\&&+~ 0.4283089945260258814058416987982628549376441370368\,\epsilon^{5}
 \nn\\&&+~ 0.1315752149611743009149867241945017928167909016942\,\epsilon^{6}
 \nn\\&&+~ \order(\!\,\epsilon^{7})
\\ I_{1011.1.2} &=& 
 +~ 0.1359328358563812618219896010403829566978061147319\,\epsilon^{4}
 \nn\\&&+~ 0.3665482868052929938485846650907631790435790974242\,\epsilon^{5}
 \nn\\&&+~ 0.3273097016981695842556322833204509862742255729220\,\epsilon^{6}
 \nn\\&&+~ \order(\!\,\epsilon^{7}) ~.
\end{eqnarray}
Our convention is that the $L$-loop master integral is normalized with respect 
to $(I_1)^L$, where the one loop vacuum bubble graph $I_1$ is defined by
\begin{equation}
I_1 ~=~ \int_k \frac{1}{[k^2+1]} 
\end{equation} 
where the propagator has unit mass. It is straightforward to restore the 
dependence on the mass $m$ in each of the above integrals using dimensional
arguments. As $I_1$ has a simple pole in $\epsilon$ then it is clear that all
these normalized master integrals are finite. However, the actual basis of 
integrals is
larger than those given in Figures $1$ and $2$. Some of the additional 
integrals at each loop order will be ultraviolet divergent as they will be 
products of $I_1$ with lower loop integrals. We have not included these as 
their values are trivial to construct.  

The next step is to determine the analytic form of these masters where we note 
that we have much higher numerical precision available than those presented 
above. To find analytic values we have used the PSLQ algorithm of
\cite{24}. Briefly the method involves trying to express the integral as a 
linear
combination of the numbers in a specific basis which a master integral is
expected
to evaluate to at each order in $\epsilon$, where the coefficients in the
combination are rationals. Once a fit has been found then it is tested against 
a higher precision numerical value of the integral to ensure the robustness of 
the final relation. Currently we have evaluated our masters to a precision of
$18000$ digits. In our case the form of the number basis is driven by 
experience with the corresponding known four dimensional vacuum bubble results,
since the latter can in principle be related to the two dimensional ones by  
dimensional recurrences, \cite{52,53}. In \cite{52,53} it was shown how to 
relate any $d$-dimensional integrals to those with the same topology in 
$(d+2)$-dimensions. 
For instance, in \cite{72} the number content of all 
possible colourings of the three loop tetrahedron topology denoted by $63$ in 
Figure $1$ was investigated in four dimensions. It was found that these 
integrals 
were related to evaluations of the Clausen function 
$\mbox{Cl}_2(\theta)$~$=$~$\sum_{n\ge1}\sin(n\theta)/n^2$~$=$~$\mbox{Im}
[\mbox{Li}_2(e^{i\theta})]$ when $\theta$ was the argument of a sixth root of 
unity such as $\theta$~$=$~$\arg{\left(\half[1+i\sqrt{3}]\right)}$. Equally 
powers of $\ln 2$ appear as well as the polylogarithm function $\mbox{Li}_n(z)$
evaluated at $z$~$=$~$\half$. 
As an example for the appearance of such numbers, let us look at 
the first non-trivial master which is the two loop sunset topology labelled $7$
in Figure $1$. It transpires that its $\epsilon$ expansion is known to all 
orders \cite{73,74} and is given by
\begin{equation}
I_7 ~=~ \frac{3}{2} \sum_{n=2}^\infty (-2)^n H_n \epsilon^n 
\end{equation} 
where
\begin{equation}
H_n ~=~ h_n ~+~ h_1 {\cal C}_{n-1}
\left( 1 - \frac{3^{\epsilon/2}\Gamma(1-\epsilon)}{\Gamma^2(1-\half\epsilon)}
\right)
\end{equation}
with
\begin{equation}
h_n ~=~ {}_{n+1}F_n \left( \frac{1}{2}, \ldots, \frac{1}{2};
\frac{3}{2}, \ldots, \frac{3}{2}; \frac{3}{4} \right) ~=~
\sum_{k=0}^\infty \frac{\Gamma(k+\half)}{(2k+1)^n \Gamma(k+1) \Gamma(\half)} 
\left( \frac{3}{4} \right)^k 
\end{equation}
and ${\cal C}_n \left( f(\epsilon) \right)$ is the coefficient of $\epsilon^n$ 
in the Taylor series of $f(\epsilon)$. For instance,
\begin{eqnarray}
H_1 &=& h_1 ~=~ \frac{2\pi}{3\sqrt{3}} ~~~,~~~ 
H_2 ~=~ h_2 ~-~ \frac{h_1}{2}\,\ln 3 ~~~,~~~ 
H_3 ~=~ h_3 ~-~ \frac{h_1}{8} \left( \ln^2 3 + 2 \zeta_2 \right) \nonumber \\
H_4 &=& h_4 ~-~ \frac{h_1}{48}  \left( \ln^3 3 + 12 \zeta_3 + 6 \zeta_2 \ln 3
\right) ~,
\end{eqnarray} 
where the above mentioned Clausen values are included as 
$9H_2$~$=$~$3\sqrt{3}\,\mbox{Cl}_2(\twopithree)$~$=$~$2\sqrt{3}\,\mbox{Cl}_2(\pithree)$.
It is straightforward to check that the numerical evaluation of the $H_n$ are in
agreement with that of $I_7$. For the higher loop integrals a second sequence 
which we found useful in our basis set of numbers was $A_n$ where
\begin{equation}
A_n ~=~ a_n ~+~ \frac{(-1)^n}{n!} \ln^n 2 
\left[ 1 - \frac{n(n-1)\zeta_2}{2\ln^2 2} \right]
\end{equation}
with
\begin{equation}
a_n ~=~ \mbox{Li}_n \left( \frac{1}{2} \right) ~=~ \sum_{k=1}^\infty 
\frac{1}{2^k k^n} ~. 
\end{equation}

Applying the PSLQ algorithm to the masters up to three loops we find 
\begin{eqnarray}
 I_{7} &=& 6 H_2 \,\epsilon^{2}-12 H_3 \,\epsilon^{3}+24 H_4\,\epsilon^{4}
-48 H_5 \,\epsilon^{5}+96 H_6\,\epsilon^{6} ~+~ \order (\epsilon^7)
\\ I_{51} &=& 7 \zeta_3 \,\epsilon^{3}+ \left[ 48 A_4-51 \zeta _4 \right] \,
\epsilon^{4}
+ \left[ 288 A_5+306 \zeta_4 \ln 2 -\frac{465 \zeta_5}{2} \right] \,
\epsilon^{5}+c_{18} \, \epsilon^{6} \nl
+~ \order (\epsilon^7)
\\ I_{62} &=& \frac{11 \zeta _3}{9} \,\epsilon^{3}+c_{1} \,\epsilon^{4}
+c_{7} \,\epsilon^{5}+c_{19} \,\epsilon^{6} ~+~ \order (\epsilon^7)
\\ I_{63} &=& \frac{\zeta_3}{3} \,\epsilon^{3}
+\left[-~8 A_4+\frac{3 c_{1}}{2}-27 H_2^2+\frac{17 \zeta _4}{2} \right] \,
\epsilon^{4} +c_{8} \,\epsilon^{5}+c_{20} \,\epsilon^{6} ~+~ 
\order (\epsilon^7) 
\end{eqnarray}
where the basis sequences $H_n$ and $A_n$ appear as well as the $\zeta_n$
series and logarithms. As noted earlier the leading order terms of these finite integrals are
not sufficient to access the coupling constant renormalization constants due
to spurious poles arising from the integration by parts. Moreover, several of
the next terms are required to resolve the same issue at four loops. In this
instance, we have included unknown coefficients $c_i$. At three loops $c_1$
appears in two of the masters but we do not need to find its analytic form as 
it turns out that it cancels with similar coefficients in various four loop 
masters when all the four loop Feynman graphs are computed for the $4$-point 
function. In fact it will become apparent below that three of the four loop 
masters contain the coefficient $c_1$ as well.

At four loops we have a larger number of undetermined coefficients but where
coefficients of $\epsilon^n$ have been found they lie within the
$\{\zeta_n, H_n, A_n \}$ basis as at three loops. Beyond the orders in 
$\epsilon$ we have computed we do not expect this basis to be complete. This
is driven by our current understanding of four loop vacuum diagrams in four
dimensions. For instance, it is known that evaluations of elliptic integrals
are present in topology $841$ \cite{42} which is structurally one of the 
simplest four
loop topologies in our master basis. The outcome of applying PSLQ at four loops
is 
\begin{eqnarray}
 I_{841} &=& c_{2} \,\epsilon^{4}+c_{9} \,\epsilon^{5}
+c_{21}\,\epsilon^{6} ~+~ \order (\epsilon^7)
\\ I_{993} &=& c_{4} \,\epsilon^{4}+c_{11} \,\epsilon^{5}+c_{23} \,\epsilon^{6} ~+~ \order (\epsilon^7)
\\ I_{952} &=& 3 \zeta_3 \,\epsilon^{4}
+[48 A_4+30\zeta_3-57\zeta _4]\,\epsilon^{5}\nl
+ \left[ 96 A_4+288 A_5+306 \zeta_4 \ln 2 -276 \zeta_3-78 \zeta_4
+\frac{39 \zeta_5}{2} \right] \,\epsilon^{6} ~+~ \order (\epsilon^7)
\\ I_{1016} &=& c_{5} \,\epsilon^{4}+c_{12} \,\epsilon^{5}
+c_{24} \,\epsilon^{6} ~+~ \order (\epsilon^7)
\\ I_{1010} &=& \left[ -\frac{c_{2}}{3}+2 c_{4}-\frac{c_{5}}{3}+5 \zeta_3
\right] \,\epsilon^{4}+c_{13} \,\epsilon^{5}+c_{25} \,\epsilon^{6} ~+~ 
\order (\epsilon^7)
\\ I_{1009} &=& \left[ \frac{2 c_{2}}{9}-\frac{10 c_{4}}{9}
-\frac{c_{5}}{6}-\frac{5 \zeta _3}{2} \right] \,\epsilon^{4}
+c_{14} \,\epsilon^{5}+c_{26} \,\epsilon^{6} ~+~ \order (\epsilon^7)
\\ I_{1020} &=& \left[ -\frac{4 c_{2}}{9}+\frac{20 c_{4}}{9}+7 \zeta_3
\right] \,\epsilon^{4} +c_{15} \,\epsilon^{5}+c_{28} \,\epsilon^{6} ~+~ 
\order (\epsilon^7)
\\ I_{1011} &=& \left[ \frac{17 c_{2}}{15}-\frac{5 c_{3}}{3}
-\frac{40 c_{4}}{9}-\frac{5 c_{5}}{9}-4 H_2-\frac{373 \zeta_3}{27} \right]
\,\epsilon^{4}+c_{16} \,\epsilon^{5}+c_{29} \,\epsilon^{6} ~+~ 
\order (\epsilon^7)
\\ I_{1022} &=& \left[ \frac{2 c_{2}}{5}-5 c_{3}+\frac{5 c_{4}}{3}
-\frac{c_{5}}{6}-12 H_2+\frac{37 \zeta_3}{18} \right] \,\epsilon^{4} \nl
+ \left[ -90 A_4+\frac{7 c_{1}}{16} -\frac{35 c_{10}}{4}-4 c_{11}
-\frac{3 c_{12}}{16}+\frac{c_{13}}{8}+\frac{57 c_{14}}{8}
+\frac{3 c_{15}}{2}-\frac{9 c_{16}}{4}+\frac{2243 c_{2}}{240} \right. \nl
\left. ~~~ +35 c_{3}
-\frac{64 c_{4}}{3}-\frac{85 c_{5}}{32}+\frac{83 c_{9}}{40}
+9 H_2^2+84 H_2+42 H_3+\frac{1663 \zeta _3}{288}+81 \zeta_4 \right] \,
\epsilon^{5} \nl +~ c_{31} \,\epsilon^{6} ~+~ \order (\epsilon^7)
\\ I_{511} &=& \left[ 4 c_{2}-20 c_{4}-2 c_{5}-\frac{176 \zeta_3}{3} \right] \,
\epsilon^{4}+c_{17} \,\epsilon^{5}+c_{32} \,\epsilon^{6} ~+~ 
\order (\epsilon^7)
\\ I_{841.1.3} &=& c_{3} \,\epsilon^{4}+c_{10} \,\epsilon^{5}+c_{22} \,
\epsilon^{6} ~+~ \order (\epsilon^7)
\\ I_{1009.1.2} &=& \left[ -\frac{77 c_{2}}{360}+\frac{25 c_{3}}{24}
+\frac{5 c_{4}}{12}+\frac{5 H_2}{2}+\frac{41 \zeta_3}{27} \right] \,
\epsilon^{4} \nl
+ \left[ \frac{8 A_4}{3}-\frac{c_{1}}{6}+\frac{25 c_{10}}{24}
+\frac{5 c_{11}}{12}-\frac{727 c_{2}}{540}-\frac{25 c_{3}}{12}
+\frac{343 c_{4}}{108}+\frac{5 c_{5}}{36}-\frac{77 c_{9}}{360}-5 H_2 \right.
\nl 
\left. ~~~ -5 H_3-\frac{5 \zeta _3}{12}-\frac{5 \zeta _4}{6} \right] \,
\epsilon^{5} +c_{27} \,\epsilon^{6} ~+~ \order (\epsilon^7)
\\ I_{1011.1.2} &=& \left[ \frac{167 c_{2}}{540}+\frac{25 c_{3}}{36}
-\frac{37 c_{4}}{18}-\frac{5 c_{5}}{27}+\frac{5 H_2}{3}
-\frac{445 \zeta_3}{81} \right] \,\epsilon^{4} \nl
+ \left[ \frac{40 A_4}{3}-\frac{5 c_{1}}{18}+\frac{65 c_{10}}{36}
+\frac{11 c_{11}}{18}-\frac{2 c_{13}}{9}-\frac{2 c_{14}}{3}
+\frac{2 c_{16}}{3}-\frac{479 c_{2}}{270}-\frac{85 c_{3}}{18}
+\frac{17 c_{4}}{6} \right. \nl
\left. ~~~ +\frac{11 c_{5}}{36}-\frac{67 c_{9}}{180}+2 H_2^2
-\frac{34 H_2}{3}-\frac{26 H_3}{3}-\frac{221 \zeta_3}{324}-\frac{71 \zeta_4}{6}
\right] \,\epsilon^{5}+c_{30} \,\epsilon^{6} \nl
+~ \order (\epsilon^7) ~.
\end{eqnarray}
As is evident from the form of these expressions with the increase in loop
order there are more masters and hence more unknown coefficients in the
$\epsilon$ expansion. While it would be interesting to know their values
explicitly the various combinations which appear at low $\epsilon$ order are
such that within the explicit renormalization to determine the naive coupling
constant renormalization constant the vast majority cancel. This was not the
case for $\beta_4(g)$, (\ref{eq:betaop4}), and we had to search for a new 
linear relation using 
PSLQ. It involves $c_{30}$ which can now be eliminated in $I_{1011.1.2}$, for
instance, since
\begin{eqnarray}
\label{eq:c30}
c_{30} &=& \frac{1024 A_{4}}{27}+\frac{448 A_{5}}{9}+\frac{2 c_{1}}{9}
-\frac{25 c_{10}}{108}-\frac{31 c_{11}}{36}-\frac{5 c_{13}}{9}
-\frac{7 c_{14}}{27}+\frac{c_{16}}{3} +\frac{13 c_{2}}{18}\nl
+~\frac{103 c_{21}}{3240} - \frac{35 c_{22}}{216} -\frac{19 c_{23}}{108}
-\frac{2 c_{25}}{9} -\frac{2 c_{26}}{3} +\frac{17 c_{27}}{9}+\frac{2 c_{29}}{3}
+\frac{85 c_{3}}{54} -\frac{178 c_{4}}{81} \nl
-~ \frac{8 c_{5}}{27}+\frac{c_{7}}{27} +\frac{427 c_{9}}{1620} + 4 H_{2}^2
-8 H_{2} H_{3}+\frac{34 H_{2}}{9} +\frac{10 H_{3}}{9} -\frac{14 H_{4}}{9} \nl 
+~\frac{476 \zeta_{4} \ln2}{9} -\frac{3988 \zeta_{3}}{81} 
-\frac{1046 \zeta_{4}}{27} +\frac{523 \zeta_{5}}{27}
\end{eqnarray}
which has been verified to $18000$ digits. It is worth stressing that such a
relation would have been hard to establish systematically without the input 
from the renormalizability of the underlying quantum field theory.

\end{document}